\documentclass[fleqn,usenatbib]{mnras}
\usepackage[T1]{fontenc}

\DeclareRobustCommand{\VAN}[3]{#2}
\let\VANthebibliography\thebibliography
\def\thebibliography{\DeclareRobustCommand{\VAN}[3]{##3}\VANthebibliography}

\usepackage{xcolor}
\usepackage{subcaption}
\usepackage{graphicx}  

\usepackage{ulem}

\newcommand{\oiii}{{[\ion{O}{iii}]}}

\newcommand{\co}{CO}
\newcommand{\he}{He}
\newcommand{\coEM}{CO}

\newcommand{\coII}{\ion{CO}{(2-1)}}
\newcommand{\coIII}{\ion{CO}{(3-2)}}
\newcommand{\csII}{\ion{CS}{(2-1)}}
\newcommand{\coI}{\ion{CO}{(1-0)}}

\newcommand{\EM}{\mathtt{EM}}

\newcommand{\mol}{\mathrm{mol}}
\newcommand{\out}{\mathrm{out}}
\newcommand{\proj}{\mathrm{proj}}

\newcommand{\hi}{\ion{H}{i}}

\newcommand{\hb}{H$\,\beta$}

\newcommand{\vel}{{\rm{v}}}
\newcommand{\Dv}{{\Delta \rm{v}}}

\newcommand{\jybeam}{{$\rm{Jy\,beam^{-1}}$}}
\newcommand{\kms}{{$\rm{km\,s^{-1}}$}}
\newcommand{\degree}{{$^{\circ}$}}
\newcommand{\ergs}{$\rm{erg\,s^{-1}}$}
\newcommand{\msun}{$\rm{M_{\odot}}$}
\newcommand{\msunyr}{$\rm{M_{\odot}\,yr^{-1}}$}
\newcommand{\jykms}{$\rm{Jy\,km\,s^{-1}}$}
\newcommand{\Kkmspc}{$\rm{K\,km\,s^{-1}\,pc^{-2}}$}

\newcommand{\rms}{$\sigma_{\rm{RMS}}$}
\newcommand{\lbol}{$L_{\mathrm{Bol}}$}
\newcommand{\lagn}{$L_{\mathrm{AGN}}$}
\newcommand{\Mmol}{$M_{\mathrm{mol}}$}
\newcommand{\fHtwo}{$f_{\rm{H}_2}$}

\newcommand{\barolo}{\sc {$^{\textsc{3d}}$barolo}}

\usepackage{graphicx}	
\usepackage{amsmath}	
\usepackage{amssymb}	

\usepackage{newtxtext,newtxmath}

\title[Cold molecular gas in NGC\,3281]{Cold molecular gas outflow encasing the ionised one in the Seyfert galaxy NGC\,3281}

\author[Bruno Dall'Agnol de Oliveira et al.]{Bruno Dall'Agnol de Oliveira$^{1}$\thanks{E-mail: bruno.ddeo@gmail.com},
        Thaisa Storchi-Bergmann$^{1}$, 
        Raffaella Morganti$^{2,3}$, 
        Rogemar A. Riffel$^{4}$,
        \and
        Venkatessh Ramakrishnan$^{5,6}$,
\\
        $^{1}$Departamento de Astronomia, Universidade Federal do Rio Grande do Sul, IF, CP 15051, 91501-970 Porto Alegre, RS, Brazil\\
        $^{2}$ASTRON, the Netherlands Institute for Radio Astronomy, Oude Hoogeveensedijk 4, 7991 PD Dwingeloo, The Netherlands\\
        $^{3}$Kapteyn Astronomical Institute, University of Groningen, Postbus 800, 9700 AV Groningen, The Netherlands\\
        $^{4}$Departamento de F\'isica, CCNE, Universidade Federal de Santa Maria, 97105-900, Santa Maria, RS, Brazil\\
        $^{5}$Finnish Centre for Astronomy with ESO, University of Turku, 20014 Turku, Finland\\
        $^{6}$Astronomy Department, Universidad de Concepci\'on, Barrio Universitario S/N, Concepci\'on 4030000, Chile\\
        }
 
\date{Accepted: 13 April 2021}

\pubyear{2022}

\begin{document}
\label{firstpage}
\pagerange{\pageref{firstpage}--\pageref{lastpage}}
\maketitle

\begin{abstract}

We present ALMA {\coII} observations of the Seyfert\,2 galaxy NGC\,3281 at $\sim$\,100\,pc spatial resolution. This galaxy was previously known to present a bi-conical ionised gas outflow extending to 2\,kpc from the nucleus. The analysis of the CO moment and channel maps, as well as kinematic modelling reveals two main components in the molecular gas: one  rotating in the galaxy plane  
and another outflowing and extending up to $\sim$\,1.8\,--\,2.6\,kpc from the nucleus,  partially encasing the ionised component. The mass of the outflowing molecular gas component is $M_{\mol,\out}$\,=\,$(2.5\pm1.6){\times}10^{6}$\,{\msun}, representing $\sim$\,1.7\,--\,2\,\% of the total molecular gas seen in emission within the inner 2.3\,kpc.
The corresponding mass outflow rate and power are $\dot{M}_{\out,\mol}$\,=\,0.12\,--\,0.72\,{\msunyr} and $\dot{E}_{\out,\mol}$\,=\,(0.045\,--\,1.6)\,${\times}\,10^{40}$\,{\ergs}, which translates to a kinetic coupling efficiency with the AGN power of only $10^{-4}$\,--\,0.02\,\%.
This value reaches up to 0.1\,\% when including both the feedback in the ionised and molecular gas, as well as considering that only part of the energy couples kinetically with the gas. 
Some of the non-rotating CO emission can also be attributed to inflow in the galaxy plane towards the nucleus. 
The similarity of the CO outflow -- encasing the ionised gas one and the X-ray emission 
-- to those seen in other sources, suggests that this may be a common property of galactic outflows.

\end{abstract}




\section{Introduction}\label{sec:intro}

The energy released by Active Galactic Nuclei (AGN) as a consequence of accretion of matter to a nuclear supermassive black hole (SMBH), couples at different degrees with the interstellar and intergalactic mediums. This feedback process has been claimed to play an important role in the evolution of the galaxies \citep[e.g.][]{fabian12,silk_mamon12}, by decreasing the galaxy star-formation rate (negative feedback) \citep[e.g.][]{cicone+14,cano-diaz+12} or even raising it (positive feedback)  \citep[e.g.][]{gallagher+19,maiolino+17}. 

Nonetheless, direct evidence of an impact at large scales -- reaching beyond the galaxy limits --  has been mostly observed in massive elliptical galaxies impacting their surrounding hot halos, via  mechanical energy released by a nuclear radio jet preventing or slowing the halo cooling rate
\cite[e.g.][]{cavagnolo+10, fabian12}. The related feedback mechanism has been referred to as radio/jet/mechanical mode, and is characterized by a low mass accretion rate to the SMBH, with Eddington ratio $\lambda_{\rm{Edd}}\lesssim$\,1\%
\citep{heckman_best14}.  

AGN feedback may also be dominated by another mechanism -- the radiative/quasar mode feedback -- in which the energy is primarily released through radiation and winds from the accretion disk  \citep{morganti17}, and the mass accretion to the SMBH occurs at a high rate, with $\lambda_{\rm{Edd}}\gtrsim$\,1\% \citep{heckman_best14}. 
However, the radiative-mode feedback has sparse evidences of an impact at large -- galactic to extragalactic scales \citep{fabian12}.

The energy released by the AGN can interact with the gas at different phases \citep{harrison18}, with the ionised phase being the most studied. However, in previous observations of outflows of ionised gas, the fraction of the AGN bolometric luminosity (\lagn) that couples kinematically with the ionised gas is usually not large enough to affect the evolution of the galaxies \citep[e.g.][]{dallagnol+2021,spence+18,husemann+16}, with typical values below 0.5\,--\,5\,\%, the minimum required by some models \citep[e.g.][]{hopkins_elvis10,zubovas18,dimatteo+05} and used in the feedback recipes of cosmological simulations \citep[e.g.][]{nelson+19,schaye+15}.
But one might be cautious when comparing observations with models/simulations \citep{harrison18}, since the percent of the total outflow energy that is kinetic can be little as  $\sim$\,20\,--\,30\,\% \citep{richings+18a}.

It is also important to consider the energy feedback other gas phases, such as the neutral atomic, the molecular and the highly ionised gas \citep{cicone+18}.
Of great importance is the molecular gas phase, specially at low temperatures ($T$ < 100\,K).
Alongside with the neutral atomic gas phase, the cold molecular gas is the fuel to form new stars \citep{veilleux+20,saintonge+22}. Any large impact on these gas phases will therefore affect the growth of the host galaxies.

One way to look for a direct global impact of AGN feedback over the cold molecular gas is to compare its content on AGN host galaxies relative to control galaxy samples. Signs of gas depletion in AGN hosts may indicate that this component has been destroyed or removed from the galaxy disk. 
Measurements of the fraction (\fHtwo) of molecular gas mass (\Mmol) relative to the total stellar mass ($M_*$), measured for the host galaxies of local AGNs shows similar values of {\fHtwo} in comparison with control samples \citep[e.g.][]{jarvis+20,rosario+18,salvestrini+22}, although there are sparse evidence of depletion in high-z objects \citep[e.g.][]{bischetti+19,circosta+21}. 
Equivalent results are found for the specific star-formation rate $sSFR$\,=\,$SFR$\,/\,$M_*$ in AGN, for which it is observed a null or positive trend between the star-formation rate ($SFR$) and {\lagn} \citep[e.g.][]{ji+22,kim+22}, which may be interpreted as the AGN not affecting the efficiency at which new stars are born. Nonetheless, \citet{ward+22} did not find a relative decrease in the {\fHtwo} and $sSFR$ of AGN in cosmological simulations, even though the inclusion of AGN feedback recipes are required to reproduce some observables such as the stellar mass function of galaxies \citep{naab_ostriker17}. 

Since a global view of the molecular content shows ambiguous results, high spatial resolution observations may shed some light on the effect of the AGN feedback on the cold molecular gas, specially if compared with other gas phases such as the ionised one. For such, we can study local AGNs, since they can be observed with high signal-to-noise ratio at such spatial resolutions, using radio interferometers in (sub)milimeter wavelengths such as the \text{Atacama Large Millimeter/submillimeter Array}. 

In this work, we analyse the molecular gas content of the local Seyfert 2 spiral galaxy NGC\,3281\footnote{See also \url{https://dc.zah.uni-heidelberg.de/sasmirala/q/prod/qp/NGC\%203281} for a comprehensive summary \citep{asmus+14}}, with bulge-to-total luminosity ratio of B/T\,$\sim$\,0.1 \citep{gao+19}. It has a Compton-thick nucleus with a column density of $\sim$\,2\,$\times$\,$10^{24}\rm{cm}^{-2}$ \citep{vignali_comastri02,sales+11}, being part of the Swift Burst Alert Telescope (BAT) catalog \citep{oh+18}, and has classified as a hidden/buried source \citep{winter+09}. 
From  the observed {\oiii} luminosity $L_{\oiii}$\,=\,$10^{40.74}$\,{\ergs} \citep{schmitt+03}, we obtain a bolometric luminosity of {\lbol}\,=\,$10^{44.3\pm0.4}$\,{\ergs} using the relations of \citet{heckman+04}. This value is compatible with the value from \citet[for the high $N_H$ case]{vasudevan+10} and \citet{stone+16}, but lower than $10^{45.28}$\,{\ergs} from \citet{sales+11}. \citet{vasudevan+10} also provide an estimate for SMBH mass of $M_{BH}$\,$\sim$\,$10^{7.2}$\,{\msun} from the K-band bulge magnitude.
The literature present a wide range of values for the total stellar mass of NGC\,3281, with values ranging between $10^{8.6}$\,{\msun} \citep[as derived from the 2MASS K-band magnitude]{winter+09} and $10^{10.24}$\,{\msun} \citep[from stellar populaton fit to the ugriz photometry]{koss+11}, with a value of $10^{9.6}$\,{\msun} from \citet[as derived from its WISE colors]{hess+15}.

We observed the {\coII} line -- which is a tracer for the cold H$_2$ \citep{bolatto+13} -- in NGC\,3283, using interferometry data from \text{ALMA} with a spatial resolution of $\sim$\,0.5\arcsec\,$\sim$\,100\,pc.
In Section\,\ref{sec:obs}, we describe the observations, alongside with ancillary data from other instruments. In Section\,\ref{sec:methodology}, the methodology and results are presented, including the identification of molecular outflows and inflows of the {\co} line.
In Section\,\ref{sec:Discussion}, we discuss the scenario of NGC\,3281, comparing it with previous works. In this section, we also obtain the molecular mass outflow and its energetics. And finally, in Section\,\ref{sec:conclusions}, we summarize the conclusions.
Throughout the paper, we use an angular scale of 0.23\,kpc\,arcsec$^{-1}$ corresponding to a luminosity distance $D_L$ of 48.54\,Mpc, obtained for a $\mathrm{H_0=70\,km\,s^{-1}}$, $\mathrm{\Omega_M=0.27}$ and $\mathrm{\Omega_\Lambda=0.7}$ cosmology, using a redshift of z\,=\,0.01124 (see Section\,\ref{sec:barolo}). Unless otherwise specified, all velocities are in the Kinematic Local Standard of Rest (LSRK), and uses a radio velocity definition.

\section{Observations}\label{sec:obs}

\subsection{ALMA data}
NGC\,3281 was observed with ALMA during 2019-04-17 in Cycle 6 (ID: 2018.1.00211.S, PI: Ramakrishnan, V.). Four spectral windows (SPW) were used, centered on the observed frequencies 228.084, 229.938, 215.389 and 217.070\,GHz. The first one contains the desired emission line of {\coII} (230.538\,GHz rest frequency), having  480 channels in a 1.875\,GHz bandwidth, and a spectral resolution (channel width) of $\Dv$\,{=}\,5.14\,{\kms}. The other three SPWs contain 128 channels with 2\,GHz bandwidth, and were used to detect the 217\,--\,233\,GHz rest frequency continuum and remove the underlying continuum emission from the emission line.

We used the default Alma Pipeline Reduction, which resulted in a continuum subtracted cube, that has a synthesized beam with full width at half maximum  of FWHM\,=\,0.57{\arcsec}{x}\,0.46{\arcsec} and Position Angle of PA\,=\,-64.4{\degree},   
and a pixel width (spatial dimension) of CDELT\,=\,0.09{\arcsec}. 
The de-convolution and imaging used Briggs weighting with a robust parameter of 0.5.
The RMS at the center of the field is 0.79\,mJy\,BEAM$^{-1}$ over the channel width. 
This is the value prior to primary beam response correction. Throughout the paper, we used a RMS 2D map, obtained by  dividing the RMS 1D value by the primary beam response image. This gives a more fiducial RMS threshold value, being 1.5 times higher than the central value at  10{\arcsec} from the nucleus, and 5 times at the end border of the field.

\subsection{Ancillary data}

\begin{figure}
    \centering
	\includegraphics[width=1.\columnwidth]{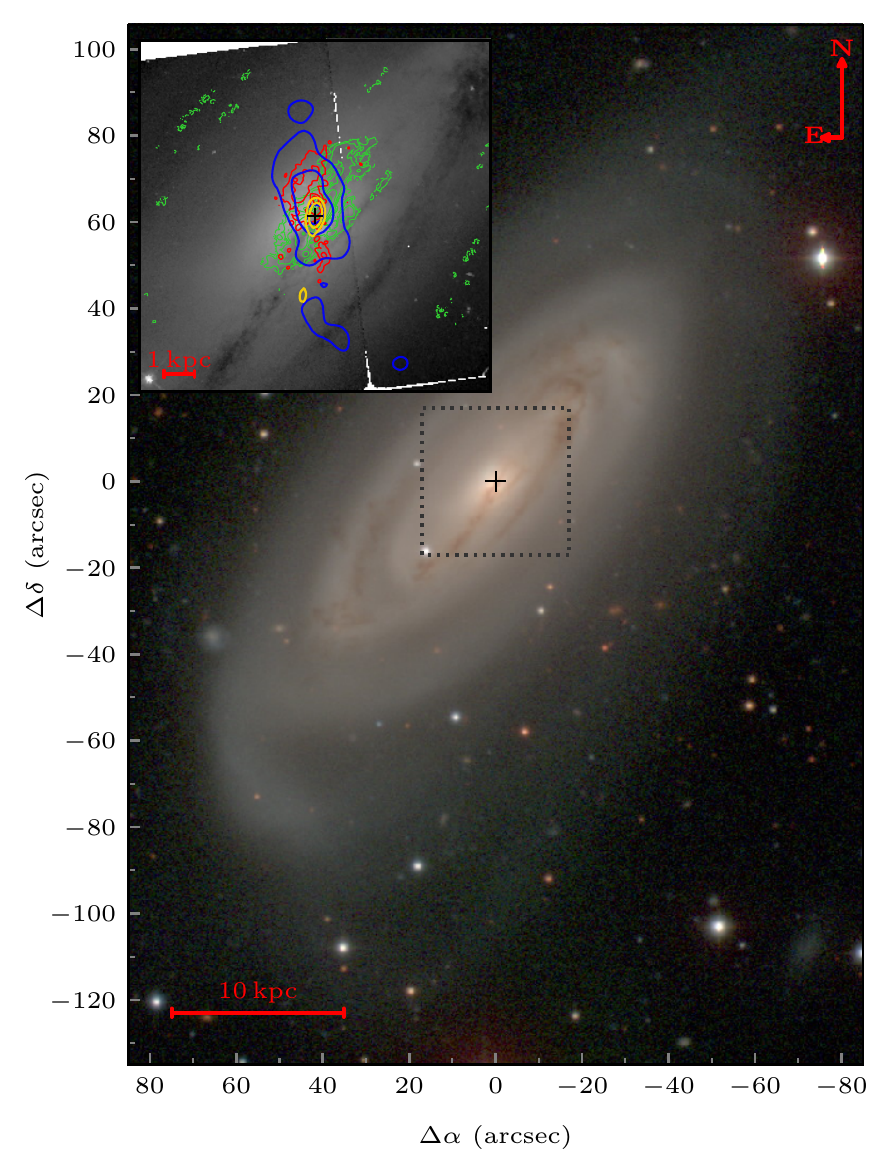}
    \caption{
    Global view of NGC\,3281. The color-composite image (from the du Pont telescope)
    shows the presence of spirals and dust.
    The zoomed-in region (delineated by the dotted box) shows the \textit{HST} continuum image obtained through the F606W filter, and includes the contours of the (\textit{HST}) {\oiii} narrow-band image (red), the Chandra's soft X-Rays 0.3\,--\,3\,keV image (blue) \citep{ma+20}, the \textit{ATCA} radio 8.6\,GHz image (yellow) \citep{morganti+99} and the $M_0$ from {\coII} (green, also shown in Fig.\,\ref{fig:map_ngc3281}). The cross marks the position of the nucleus, defined as the peak of the 230\,GHz continuum.  The outer regions of the large scale image shows that the stellar disk is warped at larger scales.
    }
    \label{fig:fig0}
\end{figure}

In Fig.\,\ref{fig:fig0}, we show the BVI color-composite image of the galaxy from \citet{ho+11}\footnote{ \url{https://cgs.obs.carnegiescience.edu/CGS/object_html_pages/NGC3281.html}} and obtained with the du Pont telescope, that reveals that the galaxy has its plane inclined by ${\sim}\,69${\degree} relative to the plane of the sky and is warped in the outer regions, with a line of nodes at PA\,${\sim}\,138${\degree} \citep{ho+11,rubin+85,storchi+92}.
The insert shows an optical continuum image obtained with the Hubble Space Telescope (hereafter \textit{HST}) through the filter F606W, with darker regions tracing dust in the galaxy.
The contours show the region where a narrow-band {\oiii}$\lambda5007$ image (in red), as well as a soft X-Rays 0.3\,--\,3\,keV one (in blue, at (10$^{-3}$, 10$^{-2.4}$, 10$^{-1.8}$)\,counts/spaxel), reveal a bipolar structure, to the north and south of the nucleus, attributed to an AGN driven outflow \citep{storchi+92}. As concluded by these authors, the NE is the far side of the galaxy and the SW is its near side, and thus, while the outflow to the north is projected against the far side of the galaxy, the southern part of the outflow is partially hidden by the near side of the galaxy plane. 
The yellow contours {purple}(at 5 and 20\,$\cdot$\,$\sigma_{\rm{RMS}\,=\,0.2\,mJy\,beam^{-1}}$) show that this galaxy has a somewhat compact radio emission at $\sim$\,1{\arcsec} spatial resolution \citep{morganti+99}, showing a second blob of 3.5\,cm (8.6\,GHz) emission at $\sim$\,8{\arcsec} approximately to the south of the nucleus, at a Position Angle of $\sim$\,170{\degree}, with a signal-to-noise ratio of $\sim$\,9. Therefore, we cannot discard the presence of a radio jet, which has been claimed to have important role in the feedback even in radio-quiet objects in quasar-mode \citep{jarvis+19}.

Fig.\,\ref{fig:fig0} also includes the flux distribution (moment $M_0$) of the {\coII} in emission as green contours (see Section\,\ref{sec:moments}),
in order to show the location of the region corresponding to the {\co} emission -- that is the main focus of the discussion in the next sections and figures -- relative to the host galaxy.

Both \textit{HST} images were retrieved from The
Mikulski Archive for Space Telescopes (MAST). The F606W continuum image proposal ID is 8598 and the galaxy was observed in 1995-04-01, while the proposal ID for which the {\oiii} image was obtained is 5479 and the observations conducted in 2001-03-03 \citep{schmitt+03}. The {\oiii} image was continuum-subtracted using an image observed as part of the same proposal. Both images have been cleaned from an excess of cosmic rays.

The soft X-ray data (events) was collected from the Chandra Data Archive, and corresponds to observations made during 2019-01-24 with ObsID 21419 \citep{ma+20}. We performed the same reduction steps described by the authors to generate an adaptively smoothed image over the  0.3\,--\,3 keV spectral range.

The 8.6\,GHz radio data was observed with the Australia Telescope Compact Array (\textit{ATCA}) during 1995-07-19 (project code C405) and the reduction process are described in \citet{morganti+99}.

\section{Methodology and results}\label{sec:methodology}

We have analysed the ALMA data using measurements of the moments and fitting a 3D model to the {\coII} emission line profiles with the {\barolo} package \citep{diteodoro_fraternali15}. We further analysed the data using channel maps of the {\coII}. 

\subsection{Moments}
\label{sec:moments}

In order to provide a global view of the distribution and kinematics of the cold molecular gas in NGC\,3281, we have obtained moment maps $M_0$, $M_1$ and $M_2$ of 
{\co}, as described in the Appendix\,\ref{sec:ap-moments}.

\begin{figure*}
	\includegraphics[width=1\linewidth]{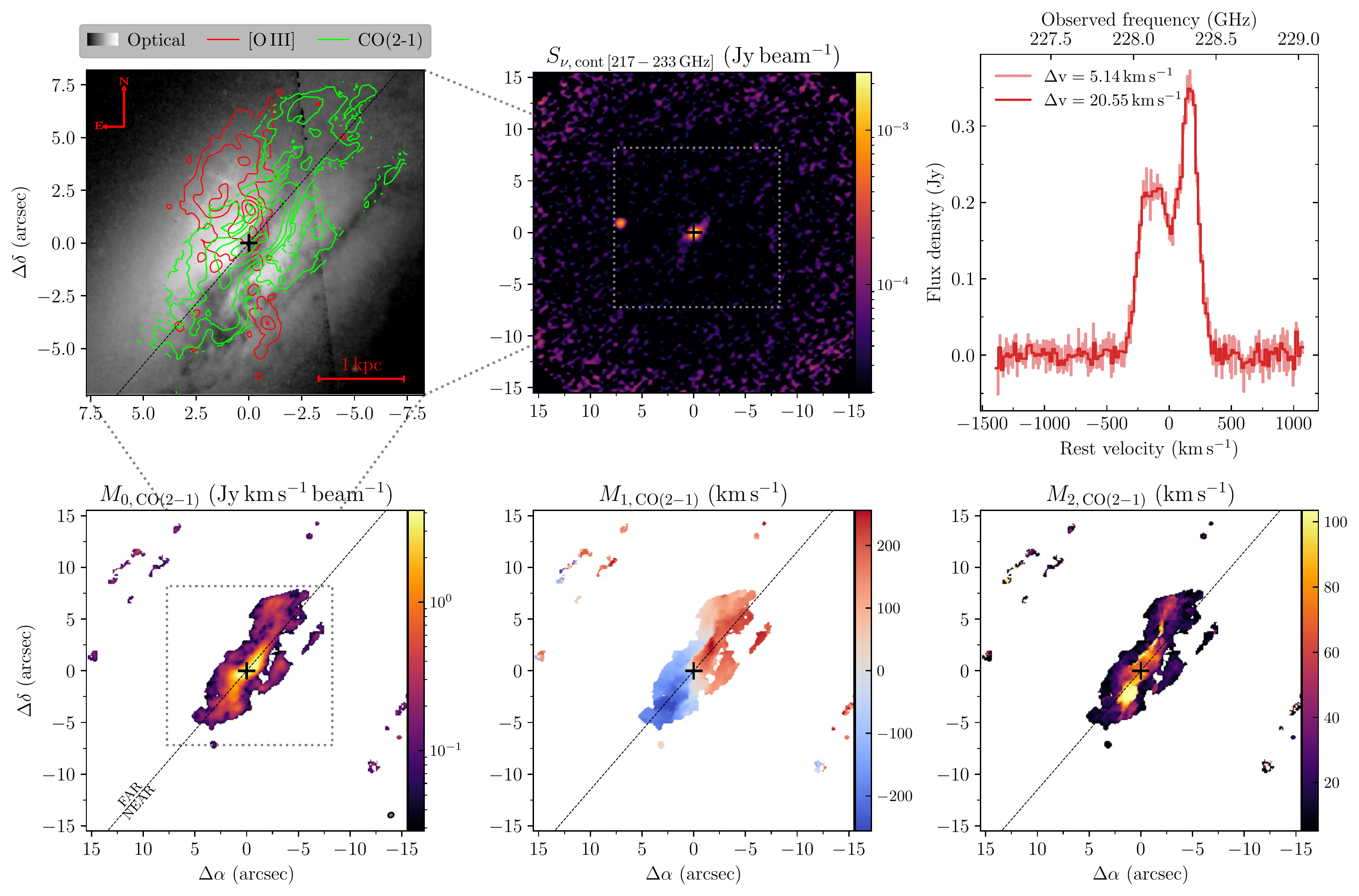}
    \caption{Optical images and CO measurements of NGC\,3281. Upper left: HST optical F606W continuum image with superposed contours of {\oiii} (red) and {\coII} (green). 
    This panel is a zoom-in of the remaining maps of this figure, as indicated by the gray dotted boxes.
    Upper center: sub-millimeter continuum image. 
    Upper right: {\co} integrated spectra, obtained from spaxels used to obtain the moments of the {\co} in emission. 
    The light colors correspond to a spectral resolution of $\Dv$\,{=}\,5.14\,{\kms}, with the darker referring to the $\Dv$\,{=}\,20.55,{\kms} re-sampled cube. 
    Lower left, center and right: the three moments of the {\co} in emission. The bottom left panel shows the synthesized beam (gray ellipse in the bottom right corner), and near/far sides of the galaxy disk plane.}
    \label{fig:map_ngc3281}
\end{figure*}

The {\coEM} moment maps of NGC\,3281 are shown in the second row of Fig.\,\ref{fig:map_ngc3281}, following the layout of a previous similar study of our group \citep{ramakrishnan+19}.
Only channels with flux densities above 3\,{\rms} were used. 
We performed the moments calculations in a cube re-binned by collapsing 4 channels into one (from channels with $\Dv$\,{=}\,5.14\,{\kms} to channels with $\Dv$\,{=}\,20.55\,{\kms}, in radio units at rest),
that reduced the noise without affecting the resulting moments values.  
We note that $M_2$ values below $\sim$\,20\,{\kms} are quite uncertain, since they originate mostly from spectra profiles for which the moments were calculated using less than 5 spectral pixels (given the 3\,{\rms} threshold). Using the original non-binned cube improved the results in only few spectra, since these are the ones with lower signal-to-noise ratio values. $M_0$ and $M_1$ are less affected by this problem.

The upper right panel of Fig.\,\ref{fig:map_ngc3281} shows the integrated spectrum in red (using only non-masked regions, see right panel in Fig.\,\ref{fig:mask_ngc3281}), both for the cube with the original channel width (lighter color) and the re-binned one (darker). 
In this case, we integrated spectra from the spaxels that were not emission dominated: 
the spaxels not masked in the {\coEM} moments maps. Only spaxels inside a 13{\arcsec} radius were considered in order to avoid spectra with lower signal-to-noise from the outer regions.

In the upper left panel of Fig. \,\ref{fig:map_ngc3281}, we added the $M_{0}$ green contours over the \textit{HST} optical continuum image, alongside with the \textit{HST} {\oiii} image red contours. 
The diagonal dashed black line marks the galaxy major axis (see Section\,\ref{sec:barolo}), where we identified  the far/near sides in the upper left panel. As expected, there is more dust blocking the stellar continuum in the near side (black lanes, see also Fig.\,\ref{fig:fig0}). 
We can also see that the flux distribution of the molecular gas  in emission (green contours) is well correlated with the dust lanes, which is expected since the two are usually seen together in galaxies \citep[e.g.][]{alves+99}, with the dust even been used as a tracer of molecular gas column density \citep{bolatto+13}. 
The dust also helps to protect the molecules from dissociation caused by the radiation field. 
In contrast, the ionised gas -- as traced by the {\oiii}  -- is not correlated with {\coEM} and shows a bipolar cone-like structure not in the disk, but at an angle to it \citep[also noted in][]{storchi+92}.
It is more visible to the North/North-East direction (above the galaxy disk), as its southern counterpart is partially hidden by the near-side of the disk, as previously mentioned.
The bipolar structure is also seen in soft-X ray \citep{ma+20}, also with the emission stronger at the far side (above the disk).

\subsubsection{Sub-milimiter continuum and possible companions}
The middle upper panel of Fig.\,\ref{fig:map_ngc3281} corresponds to the continuum between the 217\,--\,233\,GHz rest frequency, generated from the channels used to subtract the continuum from the data cube.
We see a nuclear emission ($\sim$\,0.9\arcsec radius) with a very weak linear -- or `S' shape -- emission ($\sim$\,3{\arcsec} maximum radius extent), which follows the CO emission. There is also a second blob of emission ($\sim$\,7{\arcsec} to the East, with a $\sim$\,0.7\arcsec radius). We do not find any counterpart in other wavelengths for this  localized emission, and therefore, we cannot distinguish if the origin comes from an object close to NGC\,3281, or far away from it along the line-of-sight. In the first case, it could be an object (e.g. a dwarf galaxy) that has recently passed close to the nuclear region. In this case, it could be related to the current event of nuclear accretion of gas, that may have ``turned on'' the AGN. 

We also note that, there is -- at ($\Delta\alpha,\,\Delta\delta)\,{\sim}$\,(60\arcsec, -35\arcsec) in Fig\,\ref{fig:fig0} -- a faint blob of emission in the optical, with $\sim$\,1.7\,arcsec radius, coming from another object that could also be related to the AGN activity. 
If this is the case, the apparent dust trail that seems to connect the object with the nucleus -- extending from the nucleus to the South-East, could be a trace of this interaction.
In fact, NGC\,3281 has been identified as part of the nearby Antlia cluster \cite[e.g.][]{hess+15}, with a projected distance $\sim$\,0.6{\degree}\,$\sim$\,0.5\,Mpc from the cluster's center (which we assumed to be in NGC\,3268). 
Since NGC\,3281 is in a somewhat denser environment, this may increase the chance of minor merger events, which could be the origin of the blob of emission in the optical. 

\citet{hess+15} also reports a {\hi} detection without an optical counterpart, $\sim$\,0.1{\degree}\,$\sim$\,85\,kpc to the East of NGC\,3281, which may be debris from a previous galaxy interaction. Supporting this hypothesis, from the $\sim$\,200 sources with similar characteristics -- {\hi} emission without previous optical counterpart -- \citet{haynes+11} found that 3/4 are probably associated with tidal debris.

However, none of the above hypotheses can be verified with our current data.

\begin{figure*}
	\includegraphics[width=.85\linewidth]{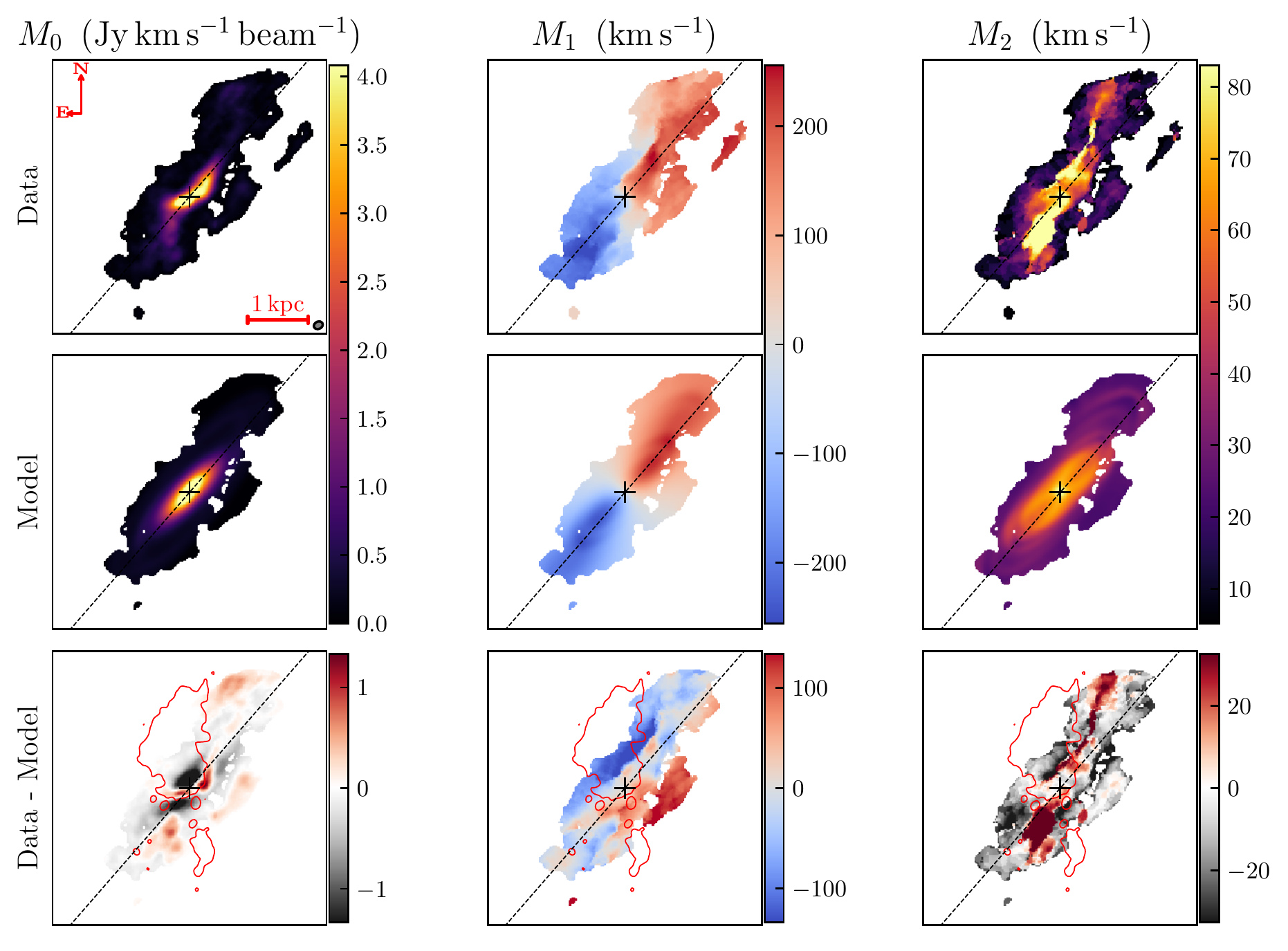}
    \caption{Fitting results of NGC\,3281 kinematics using {\barolo}. 
    The first, second and third columns show the three moments of the data (first row), the 3D model (second row), and the corresponding residuals (last row).
    }
    \label{fig:barolo_ngc3281}
\end{figure*}

\subsection{Kinematics}
\label{sec:kinematics}

\subsubsection{Tilted-rings model}
\label{sec:barolo}

We model the global kinematics of the {\coII} using the software {\barolo} \citep{diteodoro_fraternali15}. In this model's approach, tilted-ring models are fitted over the entire data cube in the spectral region covered by the selected emission line, different from the traditional 2D methods, that fit a disk model over a previously generated velocity field map (e.g. $M_1$).

With 3D models, it is possible to generate other moment maps (besides the velocity map $M_1$).
In Fig.\,\ref{fig:barolo_ngc3281} (based on one of the {\barolo} output figures), the three columns show, from left to bottom, the $M_0$, $M_1$ and $M_2$ maps of the original data cube (first row), the model (middle) and the residuals (last row). 
Fig.\,\ref{fig:barolo2D} shows the result of the fit of a 2D tilted-rings to the $M_{1}$ map \citep[from][]{begeman87}, modeled using the same software.

\begin{figure*}
	\includegraphics[width=.8\linewidth]{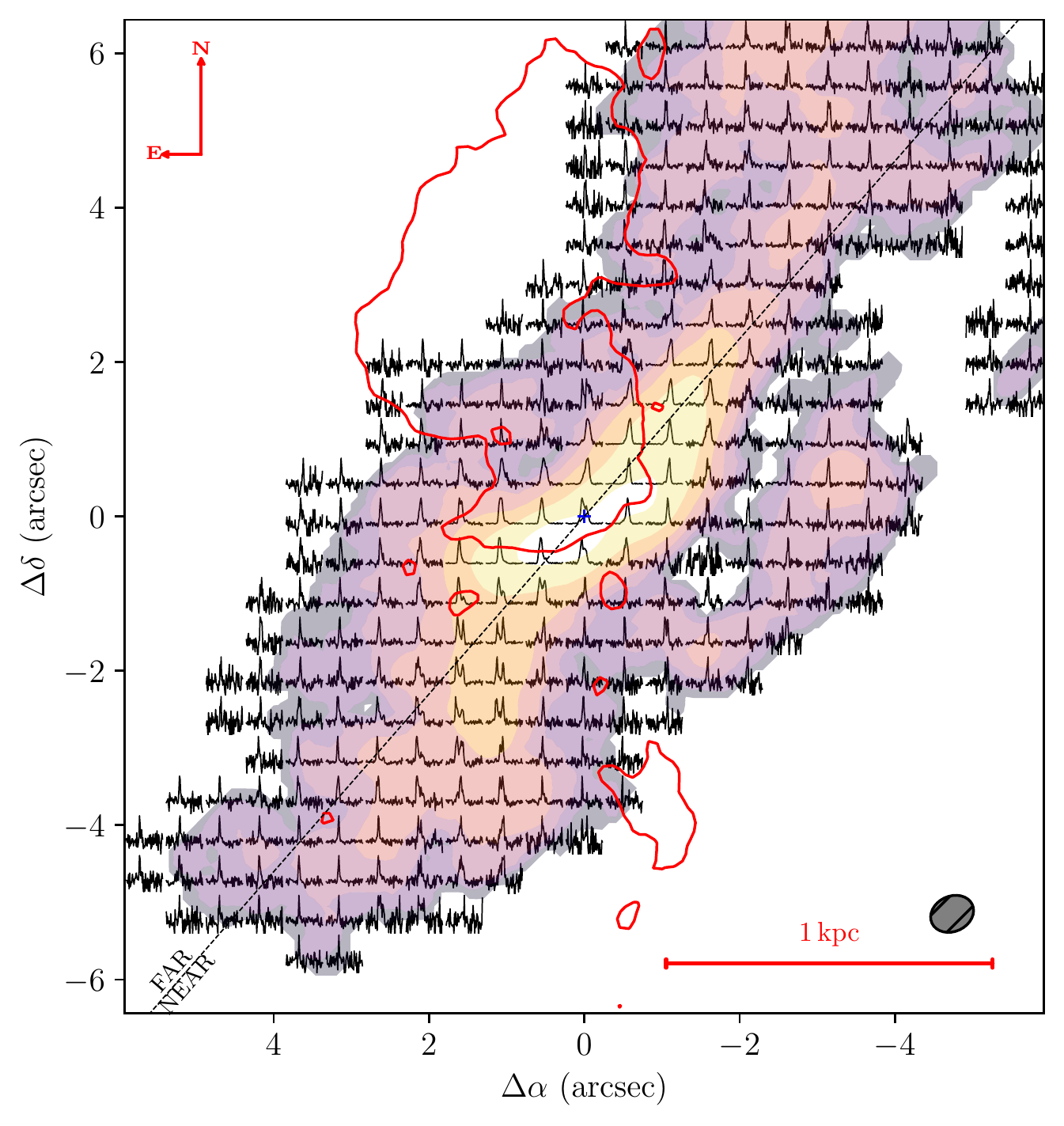}
    \caption{
    Grid of spectra from the NGC\,3281 ALMA data cube, overploted over the $M_0$. Only spaxels with emission-dominated spectra were considered. The grid has cells with width/length equal to the mean value of the beam FWHM (gray ellipse in the bottom right). Each spectrum corresponds to the sum of all spectra inside each cell. The outer contour of {\oiii} emission is shown in red.}
    \label{fig:grid}
\end{figure*}

Initial parameters used in the modelling include: 
the central position (taken from the sub-millimeter continuum), inclination $i=69${\degree} and major axis position angle $\rm{PA}=140${\degree} (the code is sensitive to these last two parameter guesses).
We adopt the following quantities as free parameters: rotational velocity, velocity dispersion, $i$ and PA (the last two being allowed to vary by $\pm5${\degree} along the radius).

We first ran the 3D fit leaving the systemic velocity as a free parameter, which resulted in the best fit value of $\vel_{\rm{sys}}$\,$\sim$\,3332\,{\kms}, in the Kinematic Local Standard of Rest (LSRK) referential frame and for a relativistic velocity definition. Using the  corresponding redshift ($z = 0.01124$), we set the data cube to the rest frame and re-fitted the data, using now a fixed $\vel_{\rm{sys}}$\,=\,0\,{\kms}, and keeping the same initial PA and $i$ values, cited above. 

From the final best fit, we obtained a mean $i$ and PA values of 139{\degree} and 73{\degree} (dashed black line in Fig.\,\ref{fig:map_ngc3281} and in the following ones), respectively. 
We note that by varying the initial values of $i$ and PA by $\pm10${\degree}, fitting results with similar residuals are obtained, which do not affect  our following analysis/conclusions below. With initial parameters outside this range, the residuals increase significantly.

Fig\,\ref{fig:barolo_ngc3281} shows that the $M_1$ -- and respective residuals -- calculated from the 3D model (second row) and the one fitted in the 2D model (last row) are quite similar.
This is expected, since, for high-resolution data (as is our case), 2D models return reliable fits to the line-of-sight field kinematics \citep{diteodoro_fraternali15}. Given this equivalence, we use only the 3D model in this paper.

The $M_{1}$ map shows small residuals along the major axis (within a projected width of $\sim$\,1.8\arcsec$\sim$\,400pc), but higher values perpendicular to it (up to $\sim\,$100\,{\kms}), in the direction of the {\oiii} emission (red contours in Fig.\ref{fig:barolo_ngc3281}). The $M_2$ and $M_0$ maps also show high discrepancies. Noticeably, $M_2$ shows positive high-residuals along a strip (redder values in the residual map), that starts parallel to the major axis, but deviates to the North in the far side, and to the South in the near side. This ``line''  separates the region where the bulk of molecular gas follows a more ordered rotational motion (along the major axis), and the perpendicular region containing the remaining molecular gas emission.
There is also a significant deviation in the $M_0$ residuals close to the nuclear region.   

\subsubsection{{\co} emission line profiles}\label{sec:profiles}

In Fig\,\ref{fig:grid}, we show a grid of {\coII} emission-line profiles overplotted on the first moment map, to identify its location over the molecular gas distribution. Each spectrum corresponds to the sum of all spectra -- that weren't masked -- inside cells with size equals to one mean FWHM.
There is a clear complexity in the profiles, including multiple components. 

For example, the strip with high $M_2$ residuals (reddest values in the lower right panel of Fig\,\ref{fig:barolo_ngc3281}) is located in a region with double-peaked profiles. 
In the Appendix\,\ref{sec:ap-moments} (Fig.\,\ref{fig:moment}), we show that double-peaked profiles can present the same moments as single-peak ones, which reveals that there is a degeneracy in the moments values for different line profiles. 
Therefore, it is important to also look at the distribution of emission-line profiles over the observed field. 
Comparing the profiles distribution with the moment maps, we note that -- along the major axis -- there is a component that dominates the emission and is tracing molecular gas with circular motion in the galaxy plane. 
However, perpendicular to it, we see a prevalence of emission line profiles with components that capture peculiar --  non-rotational motions -- in the molecular gas.
The strip with high $M_2$ residuals, therefore, shows the intersection of these two kinematically different regions, where the components have similar intensities. This also seems to be the origin of the S-shape morphology seen in the $M_0$ map, since it includes the sum of the fluxes from both components. 

\subsubsection{Channel maps}
\label{sec:channel}

\begin{figure*}
	\includegraphics[width=1\linewidth]{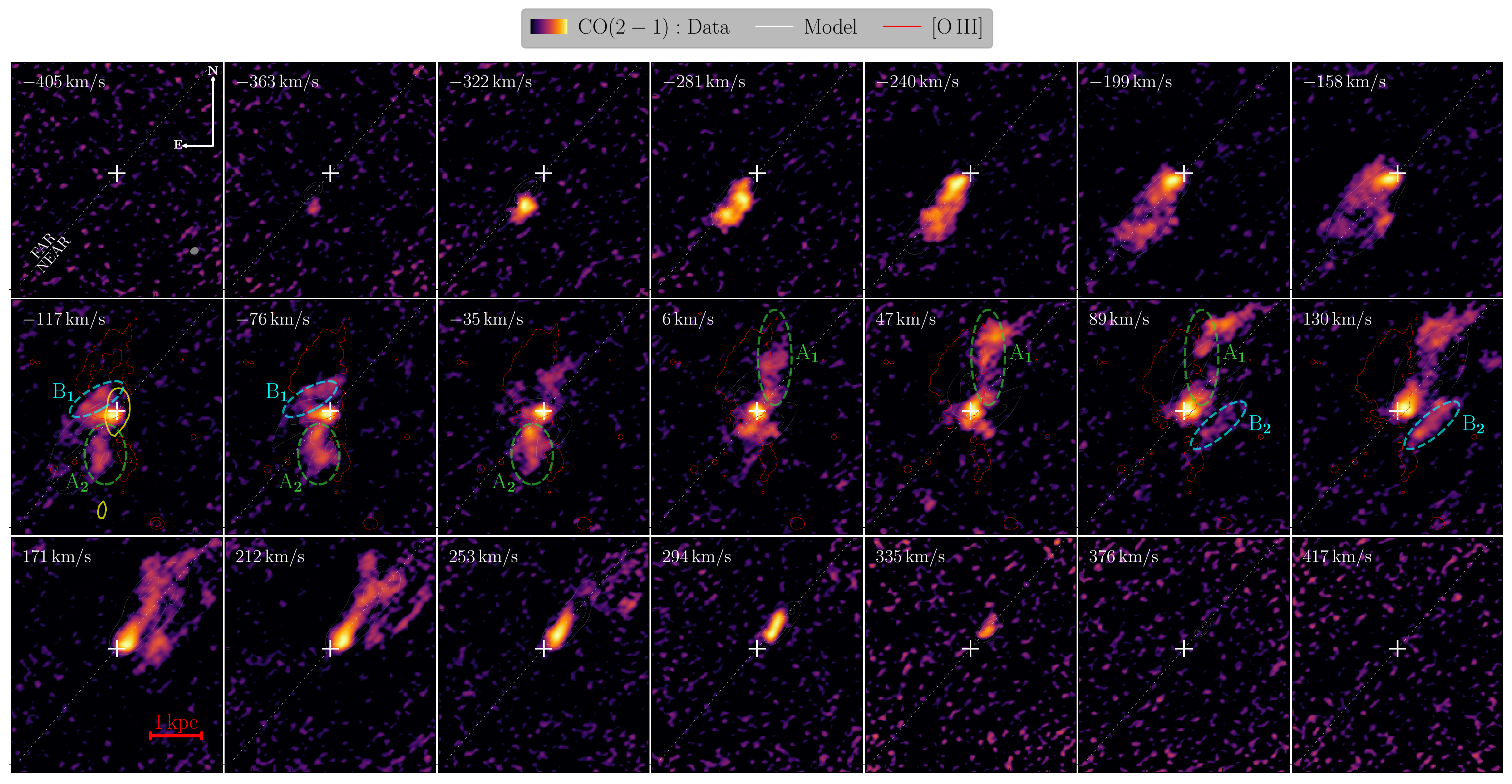}
    \caption{Channel maps of {\coII} from NGC\,3281, with the corresponding 3D model plotted as white contours. The central velocity of the integrated channels is shown in the upper right of each panel. 
    The green (A$_1$ and A$_2$) and blue (B$_1$ and B$_2$) dashed ovals indicate regions with discrepant kinematics between model and data, which we associated with molecular outflows and inflows, respectively (see the discussion in Section\,\ref{sec:channel} and the model in Fig.\,\ref{fig:ngc321_model}). The {\oiii} ionisation cone emission is shown via its outermost contour in red.
    In the $-117$\,{\kms} channel, we added the 8.6\,GHz radio in yellow contours (see also the inset in Fig.\,\ref{fig:fig0}). to show its projected alignment with the molecular outflow (A$_2$). In this channel, we also added more contours to the {\oiii}, to emphasize its non-continuity in the B$_2$ region. 
    }
    \label{fig:channel_ngc3281}
\end{figure*}

We have also obtained channel maps from the data, comparing them with the {\barolo} model in Fig.\,\ref{fig:channel_ngc3281}. 
Each panel shows the integrated flux within channels with width $\Dv$\,=\,41\,\kms velocity width, cutting values below the corresponding {\rms}. Superposed in each map are the  contours of the 3D model (in white) and of the {\oiii} line emission (red).

Once again, the channel maps show that the 3D model can reproduce the data of the {\coEM} along the kinematic major axis (white dotted line). 
However,  even along the region dominated by rotation, the fit is far from perfect, with the highest deviations seen in the channels with highest absolute velocities: above 171\,{\kms} and below $-281$\,{\kms}. 
Nonetheless, the fit is useful to identify regions that clearly deviate from disk rotation kinematics. We identify in the channel maps of Fig.\,\ref{fig:channel_ngc3281} four such regions:
A$_1$, A$_2$, B$_1$ and B$_2$, that are highlight as green and blue ellipses in the Fig.\,\ref{fig:channel_ngc3281},
and described below.

Fig.\,\ref{fig:channel_ngc3281} shows that a distinct deviation appears in the channels with small absolute velocities, specially in the  channels with central velocities between $-117$ and 89\,{\kms}, highlighted by green dashed ellipses A$_1$ and A$_2$. In these channels, we see deviations from the rings-model along the North and South directions, where there are two ``vertical'' {\co} emission structures that are parallel to the {\oiii} bipolar emission borders delineated by the red contours. 
On the far side of the galaxy, the {\co} vertical emitting structure  (A$_1$, from channels 6 to 89\,{\kms}) is observed running along the West side of the ionised gas {\oiii} emission that is above the disk. On the near side, a similar structure (A$_2$, between in channels $-117$ and $-35$\,{\kms}) is also visible to the East of the {\oiii} emission, although less pronounced, part of the reason being that the {\oiii} emission is weak, since it probably originates from behind/under the galaxy disk. 

The {\oiii} bipolar emission, which is surrounded by the CO vertical emission, has double-peaked emission line profiles with radial velocities reaching up to $\sim$\,150\,{\kms}, which are associated with an ionised outflow \citep{storchi+92}.
Therefore, one possible scenario for the A$_1$ and A$_2$ features described above is that the outflow is ``multi-phase'', where the AGN radiation that ionises the {\oiii} -- and possibly drives it away from the disk in an outflow -- also destroys most of the {\co} molecule along the bipolar region with high-excitation ionised gas. 
In this case, only at the borders of the bipolar/cone emission, the {\co} survives and is observed straddling the {\oiii} emission. 
Given that, the molecular gas in this vertical regions in the maps could also be in outflow. Since the {\coII} projected velocities are small in this region, the outflow itself has small velocities or its direction is close to the plane of the sky (inclination $i\,{\sim}$\,90{\degree}). 

\begin{figure}
	\includegraphics[width=1\linewidth]{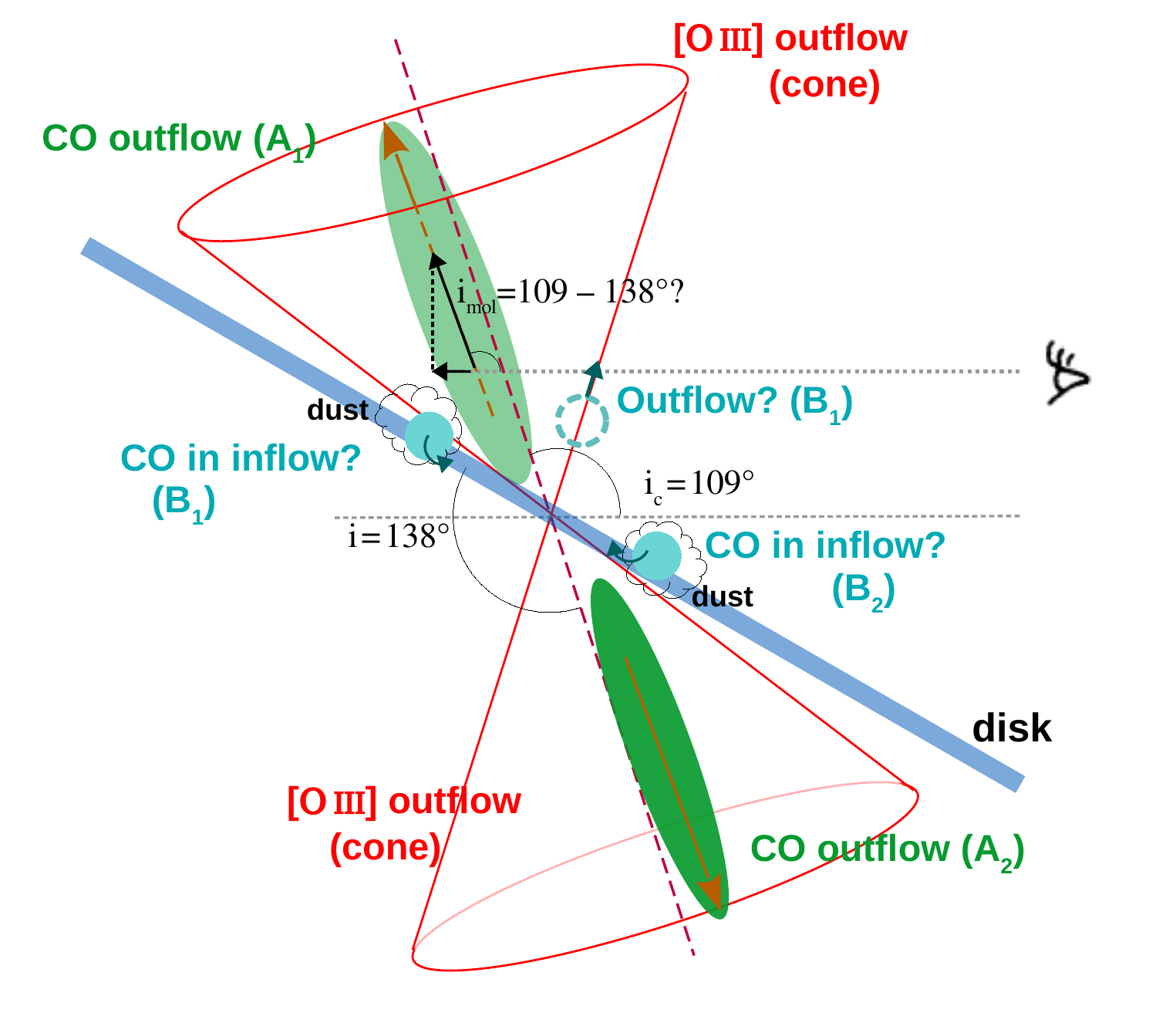}
    \caption{
    A scenario for the molecular gas outflow, based on the Model B for the ionised gas kinematics from \citet{storchi+92}, under the assumption that molecular gas outflow surrounds the ionised gas one. 
    The A$_1$ and A$_2$ green ellipses represent regions with outflowing molecular gas, that surrounds the ionised one (red cone). 
    The B$_1$ and B$_2$ blue circles identify other CO regions with non-rotating kinematics, which are elongated, but are seen from the side in this view. Both can be interpreted as molecular inflows, located close to the disk, in dust-enhanced regions. B$_2$ can also be interpreted as an outflow located in the front side of the {\oiii} cone (dashed blue circle).
    These regions are also highlighted in Fig.\,\ref{fig:channel_ngc3281}.
    }
    \label{fig:ngc321_model}
\end{figure}

This is in agreement with the model B from \citet[see their Fig.\,16b, which we adapted in Fig.\,\ref{fig:ngc321_model}]{storchi+92}, 
where the geometry of the bipolar {\oiii} outflow is a bicone with an axis with inclination $i_c\,{=}$\,109{\degree}, and the border of the cone ``touches'' the galaxy disk. In this case, the ionisation axis makes an angle $\theta_c\sim$\,20{\degree} with the galaxy disk. 
Assuming that the molecular gas outflow has the same inclination $i_{\mol}=i_c$ as that of the ionised gas (orange $A_1$ and $A_2$ ellipses in Fig.\,\ref{fig:ngc321_model}), it will reach de-projected velocities of $\sim$\,-360\,{\kms} in the near side and $\sim$270\,{\kms} in the far side.
We note, however, that there is an uncertainty in the inclination of the ionisation axis. In the case where $i_c$ has a higher value, the {\oiii} bipolar region -- depending on its opening angle -- would partly intercept the disk, and the de-projected velocities would have lower values: $\sim$\,-160\,{\kms} in the near side and $\sim$120\,{\kms} in the far side, for the limiting case of $i_{\mol}$\,=\,$i_d$=\,139{\degree}.

\begin{figure}
	\includegraphics[width=1\linewidth]{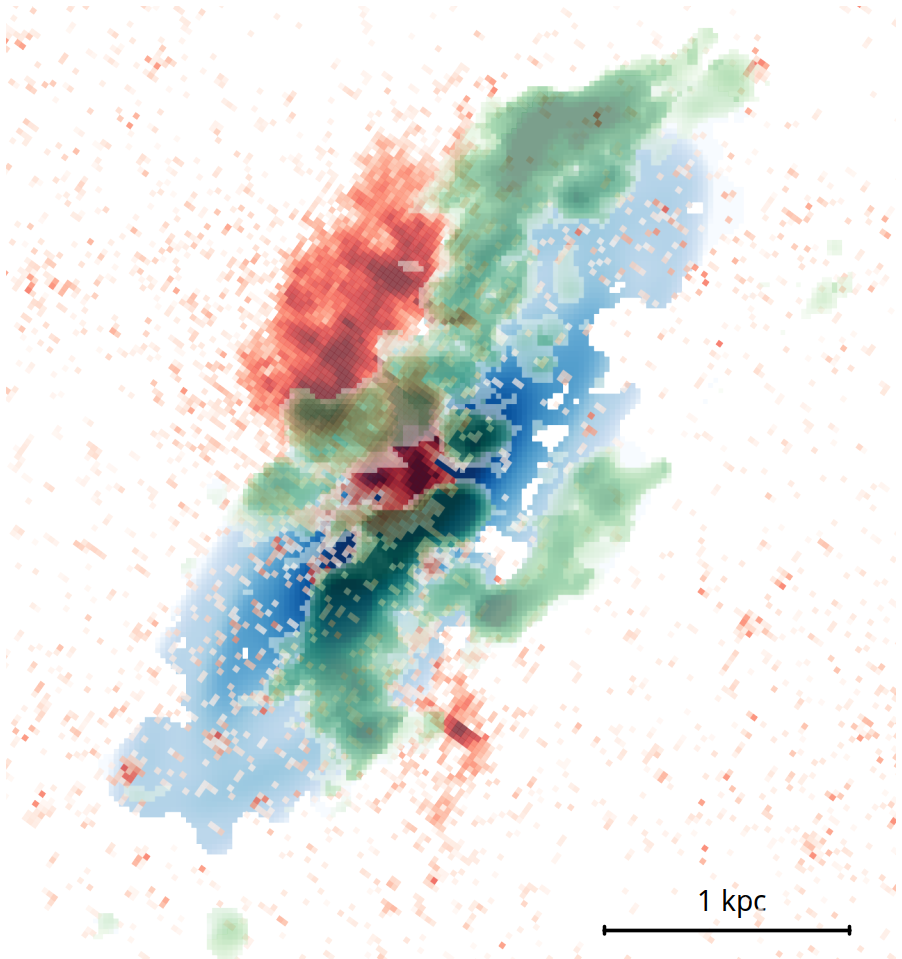}
    \caption{
    Superposition of the flux distribution maps of the {\oiii} ionised gas (red, from \textit{HST}),  
    the $M_0$ of the 3D model fitted to {\coII} data  (blue, from the middle right panel of Fig.\,\ref{fig:barolo2D}) and the residuals from this fit (green). We only integrated the residuals from channels between $-117$ and 130\,{\kms}, to highlight the A$_1$, A$_2$, B$_1$, B$_2$ regions from Fig.\,\ref{fig:channel_ngc3281}.}
    \label{fig:join}
\end{figure}

There are also some {\coII} residuals that seem to be co-spatial -- along the line-of-sight  -- with the {\oiii} emission. The blue dashed ellipses in the far side of the galaxy (B$_1$, in channels $-117$ and $-76$\,{\kms} in Fig.\,\ref{fig:channel_ngc3281}) highlights one case: a stripe of CO emission seen $\sim$\,2{\arcsec} to the north-east of the galaxy major axis, and approximately parallel to it. 
Following the hypothesis that the {\co} is destroyed by the AGN ionising radiation, the corresponding molecular gas in this region would need to be external to the ionisation cone, either behind or in front of it. 
In the model B from \citet{storchi+92}, these  CO residuals (B$_1$) would be close to the disk, if the emission origin is behind the cone  (see Fig.\,\ref{fig:ngc321_model}). 
Nevertheless, if the B$_1$ molecular gas were in outflow, 
one would expect positive velocities for this gas, but the channel maps show negative velocities. Thus, if this emission is external to the ionisation cone and in outflow, it would need to be in front of the front wall of the ionisation cone (dashed blue circle in the model of Fig.\,\ref{fig:ngc321_model}).
Another possibility is that the {\co} emission in this B$_1$ region could be due to molecular gas that is in the far side of the gas disk, but not in outflow. 
In this case, dust could be blocking the ionising radiation in that region, allowing the {\co} molecule to survive (blue circle B$_1$ in Fig.\,\ref{fig:ngc321_model}). And indeed, a close look at the {\oiii} emission distribution shows that it is not continuous, showing an apparent gap in this region. We have added more {\oiii} contour levels in the $-117$\,{\kms} channel map in Fig\,\ref{fig:channel_ngc3281} to show this gap, but a clearer view of this discontinuity in the {\oiii} flux distribution can be seen in the upper left panel of  Fig.\,\ref{fig:map_ngc3281}. 
This discontinuity in {\oiii}, ``filled'' with {\co} emission is also visible in Fig.\,\ref{fig:join} (described in the following Section\,\ref{sec:Discussion} ).

There is a somewhat similar region in the near side of the galaxy disk, to the south-west -- highlighted by the other blue ellipses in Fig.\ref{fig:channel_ngc3281} (B$_2$, from channels 89 to 130\,{\kms}). The {\co} emission from this region -- which is also roughly parallel to the major axis -- is correlated to dust lanes seen in the optical (see the darker region in the upper left panel of Fig.\,\ref{fig:map_ngc3281}). Therefore, the molecular gas in this region is probably in the galaxy disk (blue circle B$_2$ in the model of Fig.\,\ref{fig:ngc321_model}). 
The other emission seen to the north, encircled by the B$_1$ ellipses, could have a similar origin, since both are parallel to the major axis and separated from it by $\sim$\,2{\arcsec}, and there are also signs of a presence of dust in the B$_1$ region.

Given that these molecular gas B$_1$ and B$_2$ regions are not following the rotation of the galaxy disk (see Fig\,\ref{fig:barolo_ngc3281}), 
we propose that these {\coII} emitting regions are close to the disk and associated to recently accreted or disturbed molecular gas (and dust).
And, since B$_2$ is seen in redshift in the near side of the galaxy and B$_1$ in blueshift in the far side, these molecular clouds may be in inflow. 
In this case, the AGN radiation could be destroying part of the inflowing {\co}, ejecting it from the disk, which is what we see in the A$_1$ and A$_2$ ellipses in Fig.\,\ref{fig:channel_ngc3281}: a scenario of feeding and negative feedback -- via removal of part of the molecular gas reservoir. However, given the apparent low velocities, this gas may return to the disk in the future.

We note that there is also the possibility that the  outflowing  {\co} emission comes from molecular gas that cooled/condensed from the ionised gas in the cones, corresponding to ``in-situ'' formation of molecular clouds \citep{veilleux+20}. This is a proposed mechanism to account for the presence of molecular gas in winds \citep[e.g][]{silich+03}, and could be the origin of the gas that collapsed to form stars inside some galactic outflows, comprising the so-called ``positive feedback'' \cite[e.g.][]{gallagher+19,maiolino+17}. 

\section{Discussion}
\label{sec:Discussion}

In order to have a global view of the gas kinematics, we generated the Fig.\,\ref{fig:join}. 
In green, we added the flux distribution map of the {\co} residuals relative to rotation (from Fig.\,\ref{fig:channel_ngc3281}). We integrated between $-117$ and 130\,{\kms}, to include only the channels where the A$_1$, A$_2$, B$_1$ and B$_2$ are present.
In blue, we superposed the the $M_0$ obtained from the {\barolo} fit (middle left panel in Fig.\,\ref{fig:barolo_ngc3281}), that highlights the regions dominated by rotation.
And in red, we included the {\oiii} ionised gas map, from the \text{HST} narrow-band image.

This figure shows that the gas that is not rotating is partially encasing the ionised gas above the galaxy plane as well as that below the plane, suggesting a two phase outflow, present in both the ionised and in the molecular gas. 
But note that not all residuals in green are necessary in outflow. Specially, the gas inside the B$_1$ and B$_2$ regions (see Fig.\,\ref{fig:channel_ngc3281}) may have other kinematics: inflows to the nuclear region, for example.
The identification and analysis of the  A$_1$, A$_2$, B$_1$ and B$_2$ regions were done in Section\,\ref{sec:channel}.

\subsection{Scenario}

Fig.\,\ref{fig:ngc321_model} combined with Fig.\,\ref{fig:join} illustrate our proposed scenario for NGC\,3281. Fig.\,\ref{fig:ngc321_model} is based on one of the models proposed for the {\oiii} kinematics in  by \citet{storchi+92}. Our observations and measurements, including the complex double-peaked emission line profiles, as discussed in the previous sections reveal the presence of more than one kinematic component in the {\coEM} velocity field: part of the {\coII} molecular gas is in rotation in the galaxy plane, but part of it shows a contrasting kinematics, that can be interpreted as due to outflows out of the galaxy plane and also possibly inflows in the plane.

The presence of a molecular outflow, surrounding the ionised gas is supported by the flux distribution of the {\coEM}, that curves upwards and downwards from the galaxy plane: A$_1$ and A$_2$ regions in Figs.\,\ref{fig:channel_ngc3281} and \ref{fig:ngc321_model}, as well as by Fig.\,\ref{fig:join} that shows the non-rotating component of the gas kinematics surrounding the ionised gas component.
The molecular gas may have been ejected from the disk, surviving the ionising AGN radiation at the boundaries of the {\oiii} ionisation cones. 

There is also {\coEM} gas that is apparently in the galaxy plane, but not following the rotating kinematics seen along the major axis (B$_1$ and B$_2$ regions in Figs.\,\ref{fig:channel_ngc3281}). These regions run approximately parallel to the major axis, with the {\co} in the near side  of the disk (region B$_2$) lying close to a dust lane. Given the redshift/blueshift velocities in the near/far side, the emission from these regions can be interpreted as being from molecular gas in inflow. The {\co} in region B$_1$ could also be interpreted as originating in ejected clouds in the front wall of the ionised cone, but given its similarities with B$_2$ region -- including similar distances to the nucleus -- the inflow hypothesis is more compelling. 

Although the extended radio emission from NGC\,3281 is weak, the southern radio blob (see Fig.\,\ref{fig:fig0}) is located along the direction of the {\coEM} outflow {A$_2$} region (see green contours in the $-117$\,{\kms} channel in Fig.\,\ref{fig:channel_ngc3281}). Therefore, we cannot discard that the molecular outflow seen NGC\,3281 is partly driven by shocks generated by the radio-jet. 

Another interesting feature is the apparent alignment between the extended {\oiii} and soft X-ray emission. 
It turns out that it is not uncommon to find spatial correlations between the extended narrow line region (ENRL) and the soft X-Ray, especially in Compton Thick objects \citep{fabbiano_elvis22} like NGC 3281, but also in other objects such as  radio galaxies \citep{balmaverde+12}. 
The origin of extended soft X-ray has been attribute to highly ionised emission lines (e.g. \ion{O}{vii}, \ion{O}{viii}\,Ly$\alpha$, \ion{Ne}{IX}),  generated by AGN photoionisation or shocks.

\subsection{Molecular gas mass}

We now use the CO emission inside the A$_1$ and A$_2$ regions (green ellipses in  Fig.\,\ref{fig:channel_ngc3281}) to calculate the total outflowing molecular gas mass.
The total flux in emission was integrated inside a 10{\arcsec} ($\sim$\,2.3\,kpc) radius, to avoid adding the contribution from the CO emission  observed beyond this radius, which may be artifacts of the data reduction (see lower panels of Fig.\,\ref{fig:map_ngc3281}). Including it would increase that total mass by $\sim$\,6\,\%.

The total flux emitted inside these outflowing regions is $(S_{\nu}\Delta{\rm{v}})_{\out}$\,$\sim$\,1.6\,{\jykms}, which translates to a {\coII} luminosity of  $L'_{{\coII},\out}$\,${\sim}\,\,2.3\,{\times}\,10^{6}$\,{\Kkmspc} \citep{solomon+97}. 
Following \citet{ramos-almeida+22}, for a conversion factor of $\alpha_{\co}\,{=}\,0.8\pm0.5$\,{\msun}(\Kkmspc)$^{-1}$ \citep{morganti+15} and a ratio $L'_{\coII}$/$L'_{\coI}$\,=\,1  \citep{carilli+13,braine_combes92}, we obtain a total molecular mass in outflow of $M_{\mol,\out}$\,=\,$(2.5\pm1.6){\times}10^{6}$\,{\msun}, which is  distributed equally between regions A$_1$ and A$_2$.
As a comparison, the total molecular gas mass in emission is $M_{\mol,\EM}$\,=\,($1.5\pm0.9){\times}10^{8}$\,{\msun} ($S_{\nu}\Delta{\rm{v}}\,{\sim}\,95.3$\,{\jykms}).
Therefore, the corresponding fraction is $M_{\mol,\out}{/}M_{\mol,\EM}$\,$\sim$\,1.7\,\%, where we are assuming the same $\alpha_{\co}$ value for the $M_{\mol,\out}$ and $M_{\mol,\EM}$. 
If we assumed that gas inside the B$_1$ region was also outflowing, this
fraction would be $\sim$\,2\,\%. The molecular mass values were multiplied by a factor of 1.36 to account for the presence of {\he} and heavier elements \citep{saintonge+22}.

Using the compilation made by \citet{ramos-almeida+22}, we note that the  NGC\,3281 fraction of molecular gas mass in outflow is higher than the values found by the authors for their QSOs \citep[in the range 0.1\,--\,1\,\%,][]{ramos-almeida+22} and for other Seyfert galaxies \citep[0.2\,--\,0.7\,\%, ][]{alonso-herrero+19, dominguez-fernandez+20, garcia-bernete+21}.  
Aside from that, the  $M_{\mol,\out}{/}M_{\mol,\EM}$ values from NGC\,3281 are similar to those found in jetted Seyferts \citep[3\,--\,5\,\%, ][]{garcia-burillo+14, morganti+15} and ULIRG AGNs \citep[2\,--\,27\,\%, ][]{feruglio+10, cicone+14}.

\subsection{Molecular mass outflow rate and power}

To evaluate the strength of the molecular outflow, we measured the mass outflow rate ($\dot{M}_{\out,\mol}$) and outflow power ($\dot{E}_{\out,\mol}$). The contribution from each velocity channel was calculated separately, and resulting values integrated.
We applied two methods over the regions delimited by the A$_1$ and A$_2$ ellipses in  Fig.\,\ref{fig:channel_ngc3281}, using channels with the projected velocities (6, 47, 89)\,{\kms} and ($-117$, $-76$, $-35$)\,\kms, respectively.

\subsubsection{Method 1: average}\label{sec:met1}

The first method assumes that the rates do not vary along the radii, with the resulting values corresponding to averages of these quantities. 
For this, we consider a fixed outflow radius $R_\out(\vel_i)$ for each channel, with the respective outflow mass $M_{\mol,\out}(\vel_i)$ being integrated inside each of the elliptical regions. Only spaxels with flux density above {\rms} are considered. The corresponding formulas are: 

\begin{eqnarray}
    \dot{M}_{\out,\mol} = \sum_{\vel_i} \frac{\vel_i \cdot M_{\mol,\out}(\vel_i) }{R_{\out}(\vel_i)} \\
    \dot{E}_{\out,\mol} = \sum_{\vel_i} \frac{1}{2} \cdot \vel_i^2 \dot{M}_{\mol,\out}(\vel_i). 
\end{eqnarray}
In order to avoid overestimation of the sizes due to spurious spaxels, we first convolved the map of each channel by a beam-shape 2D Gaussian. We then measured the distance  along each ellipse from the closest to farthest spaxel with flux density above {\rms}. The measured projected distances ($R_\out(\vel_i)$) obtained for the channels of the A$_1$ region were (6.3, 6.75, 6.9){\arcsec}, and (4.5, 4.4, 4.3){\arcsec} for the A$_2$ region.

To correct for projection effects, we considered two inclinations $i_{\mol}$ for the molecular outflow: one equal to the inclination of ionisation axis $i_c$\,{=}\,109{\degree} \cite[from][]{storchi+92} and another equal to the inclination of the disk $i_d$\,{=}\,139{\degree}, for the case of the molecular outflow axis being closer to the disk. We corrected the projected  velocities ($\vel_{\proj}$) and radii/sizes ($r_{\proj}$) using: $\vel = \vel_{\proj} / |\cos(i_\mol)$| and $r = r_{\proj} / \sin(i_\mol)$.

For an inclination of  $i_{\mol}$\,{=}\,139{\degree}, we obtained values of $\dot{M}_{\out,\mol}^{\rm{avg}}$\,=\,$0.12\pm0.04$\,{\msunyr} and $\dot{E}_{\out,\mol}^{\rm{avg}}$\,=\,$(4.8\pm1.7)\,{\times}\,10^{38}$\,{\ergs}. 
And for an inclination of  $i_{\mol}$\,{=}\,109{\degree}, we obtained values of $\dot{M}_{\out,\mol}^{\rm{avg}}$\,=\,$0.39\pm0.12$\,{\msunyr} and $\dot{E}_{\out,\mol}^{\rm{avg}}$\,=\,$(8.1\pm2.9)\,{\times}\,10^{39}$\,{\ergs}. 
Therefore, depending on the inclination, the values of $\dot{M}_{\out,\mol}$ and $\dot{E}_{\out,\mol}$ differ by a factor of $\sim$\,3 and $\sim$\,17, respectively, with the factors being equal for the following method 2. 
The uncertainty are the result of propagating the uncertainty from $\alpha_{\co}$.

\subsubsection{Method 2: radial profiles}\label{sec:met2}

\begin{figure}
	\includegraphics[width=1\linewidth]{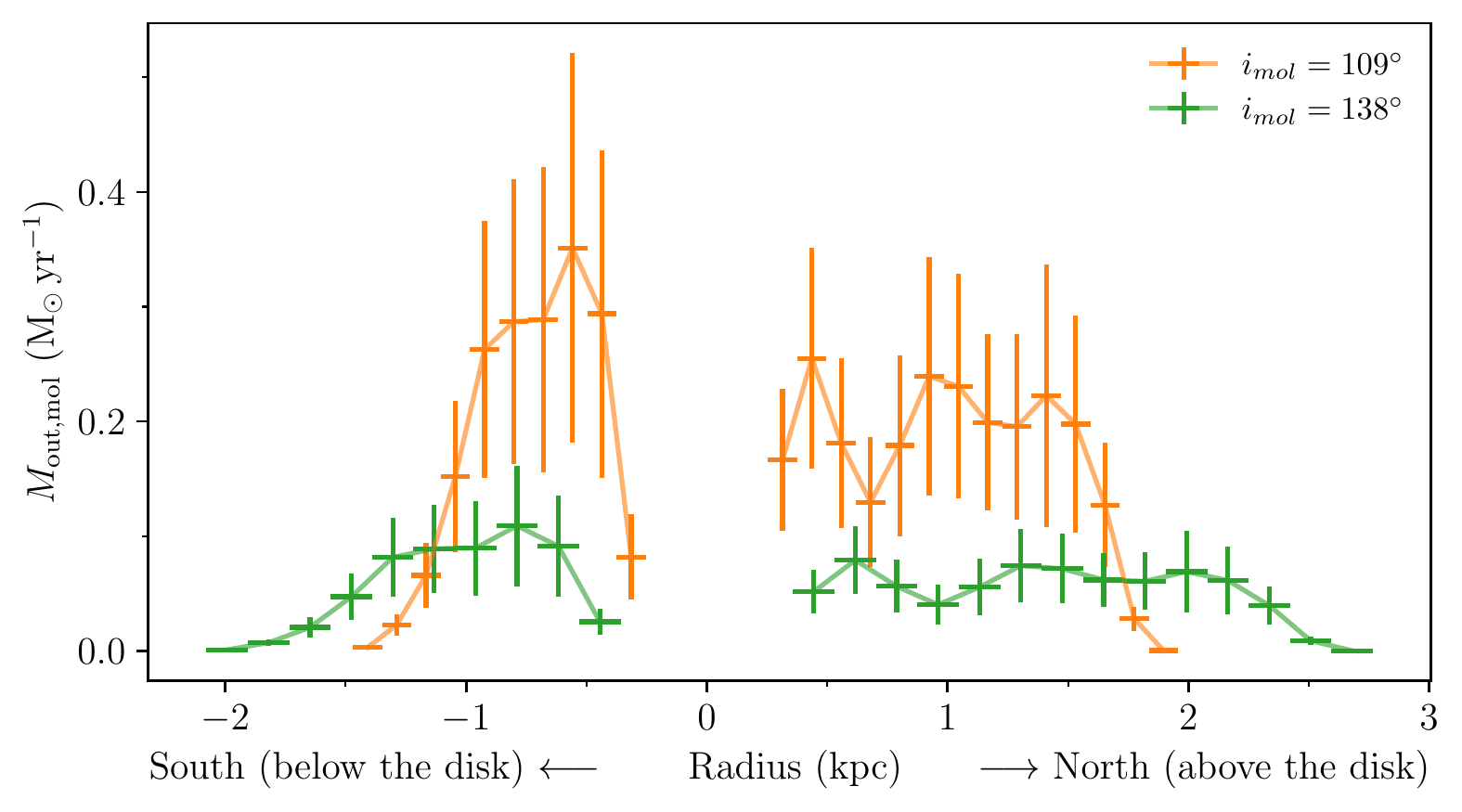}
    \caption{
    Radial profiles of the molecular mass outflow rate $M_{\out,\mol}$ from the nucleus of NGC\,3281 
    as a function of the de-projected radius. $M_{\out,\mol}$ values in orange and green correspond to outflow inclinations $i_{\mol}$ of 109{\degree} and 139{\degree}, respectively.
    }
    \label{fig:mdot}
\end{figure}

The second method, is based in generating the radial of profiles of $\dot{M}_{\out,\mol}(r)$ and $\dot{E}_{\out,\mol}(r)$, as shown in Fig.\,\ref{fig:mdot} for the first quantity. 
For each ring at radius $r$, with a defined thickness $\delta r$\,=\,FWHM$_{\rm{mean}}$\,=\,$0.5${\arcsec}, we calculate:

\begin{equation}
    \dot{M}_{\out,\mol}(r) = \sum_{\vel_i} \frac{\vel_i \cdot M_{\mol,\out}(\vel_i) }{\delta r},
\end{equation}
and analogous for $\dot{E}_{\out,\mol}(r)$. In this case, we integrated the contribution from channels inside A$_1$ (North, above the disk) and A$_2$ (South, below the disk), separately. 

Above the disk (North), where the {\coII} emission is less absorbed by the gas/dust in the disk, the $\dot{M}_{\out,\mol}(r)$ is somewhat constant up to a radius of $\sim$\,1.5\,--\,2.2\,kpc (depending on the inclination $i_\mol$).
Below the disk, it reaches a peak value at $\sim$\,0.5\,--\,0.8\,kpc, and is approximately constant up to $\sim$\,0.9\,--\,1.3\,kpc.
The maximum outflow extent is $R_{\out}^{\rm{max}}$\,$\sim$\,7.5{\arcsec}\,$\sim$\,1.8\,--\,2.6\,kpc.

To compare our data with results from the literature and the method 1 above (see Section\,\ref{sec:met1}), we measured the peak values of $\dot{M}_{\out,\mol}$ and $\dot{E}_{\out,\mol}$ \citep[e.g][]{dallagnol+2021,trindade-falcao+21}. For this, we added the contribution from each side of the cone (North and South), such that the following resulting values correspond to the maximum rates of mass crossing rings/spheres with a fixed $\delta r$. 
We obtained maximum values of $\dot{M}_{\out,\mol}^{\rm{max}}$\,=\,$0.55\pm0.17${\msunyr} and $\dot{E}_{\out,\mol}^{\rm{max}}$\,=\,$1.2\pm0.5\,{\times}\,10^{40}$\,{\ergs} for $i_{\mol}$\,{=}\,109{\degree}, at a radius of 0.4\,--\,0.6 kpc. 

Considering mass outflow rates obtained from both inclinations, the maximum range obtained from the method 2 (peak of the radial profiles) is $\dot{M}_{\out,\mol}^{\rm{max}}$\,=\,0.12\,--\,0.72\,{\msunyr}, which is in agreement with the results from the method 1 (the average values)  $\dot{M}_{\out,\mol}^{\rm{avg}}$\,=\,0.08\,--\,0.5\,{\msunyr}. The same is true for molecular outflow power, with ranges of  $\dot{E}_{\out,\mol}^{\rm{max}}$\,=\,(0.045\,--\,1.6)\,${\times}\,10^{40}$\,{\ergs} and $\dot{E}_{\out,\mol}^{\rm{avg}}$\,=\,(0.031\,--\,1.1)\,${\times}\,10^{40}$\,{\ergs}. 
Therefore, we chose to use $\dot{M}_{\out,\mol}^{\rm{max}}$ and $\dot{E}_{\out,\mol}^{\rm{max}}$ as the default values in the remaining of the paper. 
From the compilation made by \citet{ramos-almeida+22}, the mass outflow rate of NGC\,3281 falls at the low end of other Seyferts ($\sim$\,0.3\,--\,5\,{\msunyr}), with more energetic sources like QSOs and ULIRGs having higher values of $\dot{M}_{\out,\mol}$,  by up to $\sim$\,3 orders of magnitude.

To include the contribution from the ionised gas, we calculate the ionised outflow power from the data collected from \citet{storchi+92}. We used a total ionised gas mass (from {\hb}) of $M_{\rm{ion}}$\,$\sim$\,2\,{$\times$}\,$10^{6}$\,{\msun}, a {\oiii} de-projected ouflow radial velocity and velocity dispersion of $\vel_{\out,\rm{ion}}$\,$\sim$\,150\,{\kms} and $\sigma_{\out,\rm{ion}}$\,$\sim$\,150\,{\kms}, and a de-projected ionised outflow extent of $R_{\out,\rm{ion}}$\,$\sim$\,2\,kpc. Since the available data are not separated by channels (as the molecular data from our work), we calculated the outflow power using: $\dot{E}_{\out}=0.5 \cdot (\vel^2 + \sigma^2) \cdot \vel \cdot M / R$, for the conservative case of B\,=\,1 \citep{harrison18}. The resulting ionised outflow power is $\dot{E}_{\out,\rm{ion}}\,{\sim}\,10^{39}$\,{\ergs}. 
Therefore, considering the uncertainties, we cannot distinguish if the outflow power is stronger in the molecular or the ionised phase.

\subsubsection{Coupling efficiency}\label{sec:eff}

To gauge the impact that the molecular outflow may have on the evolution of NGC\,3281, we calculated the percent of AGN luminosity ({\lagn}) that couples kinetically with the molecular gas via outflows: the kinetic coupling efficiency $\varepsilon_{f,\mol}= \dot{E}_{\out,\mol} / L_{\rm{AGN}}$. 
For this, we used a luminosity of {\lbol}\,=\,$10^{44.3\pm0.4}$\,{\ergs} (see Section\,\ref{sec:intro}), and since the method 1 and 2 returned similar values (Section\,\ref{sec:met1} and \ref{sec:met2}), we used the peak value from the radial profile $\dot{E}_{\out,\mol}$ as the kinetic power.

Therefore, the resulting  $\varepsilon_{f,\mol}$ of the outflowing molecular gas ranges between $10^{-4}$ and 0.02\,\%. This value is, at least, one order of magnitude lower than the $\sim$\,0.5\,\% value required in models \citep[e.g.][]{hopkins_elvis10,zubovas18} to produce a significant impact in the galaxy and influence its evolution. Adding the contribution from  $\dot{E}_{\out,\rm{ion}}$ and assuming that only 20\% of energy couples kinematically
 with the gas \citep{richings+18a}, the maximum value $\varepsilon_{f}$ is still only 0.1\,\%.

Hence, given the current energy involved in the AGN feedback of NGC\,3281, we do not expect that the evolution of host galaxy to be affected globally over a short period of time. But we note that, since the molecular gas is being expelled from the inner regions, the star formation in the nuclei will be at least partially affected in a short period of time. This is in agreement with the results found by  \citet{storchi+92} from the equivalent width of optical stellar absorption lines. The values indicate a dominant old stellar population within 2.3 kpc of the nucleus, typical of the bulges of early-type galaxies, even tough NGC\,3281 morphology is dominated by the disk, with a bulge-to-total flux of B\,/\,T\,$\sim$\,0.1 \citep{gao+19}. 
n contrast, \citet{stone+16} obtained a SFR of $\sim$\,7\,{\msunyr} in NGC\,3281, from the luminosity at 160\,$\rm{\mu m}$.
This value would put NGC\,3281 above the star formation main sequence \citep[e.g.][]{elbaz+07}, given its total stellar mass of $10^{8.6}$\,--\,$10^{10.24}$\,{\msun} (see Section\,\ref{sec:intro}).
Considering only the molecular gas outflow, the corresponding mass loading factor $\eta_{\mol}$\,=\, $\dot{M}_{\out,\mol}$/SFR\, $\sim$\,0.02\,--\,0.1 reinforces the conclusion that the AGN feedback in NGC\,3281 is not strongly regulating the star formation, given that galaxy evolution models usually require $\eta$\,$\sim$\,1 \citep{veilleux+20}.

\subsection{Comparison with previous studies}

A similar molecular gas outflow -- surrounding an ionised outflow -- has been observed in {\coI} in the starburst galaxy NGC\,253 \citep{bolatto+13b}, where the corresponding $\dot{M}_{\out,\mol}$ is 3\,--\,9\,{\msunyr}, an order of magnitude higher than our value for NGC\,3281.
This galaxy has a total stellar mass of $M_*$\,${\sim}\,10^{10.5}$\,{\msun} \citep{,bailin+11,lucero+15}, while NGC\,3281 apparently has a lower stellar mass, with a value ranging between $10^{8.6}$\,--\,$10^{10.24}$\,{\msun} (see Section\,\ref{sec:intro}).
Therefore, the difference in the host $M_*$ may affect a direct comparison between the two galaxies, with the same being true for the other galaxies discussed below.

In our own galaxy, the Milk Way ($M_*$\,${\sim}\,10^{10.8}$\,{\msun}, from \citet{mcmillan+11}), 
the molecular line {\csII} emission map displays an ``hourglass'' morphology \citep{hsieh+16}, surrounding the base of a cavity -- filled with X-ray/radio lobes  (similar to our Fig.\,\ref{fig:fig0}) \citep{ponti+19,heywood+19,bland-hawthorn+03} -- that may have been caused by a previous outflow \citep{veilleux+20}.   The scales involved, however, are much smaller in the Milky Way, that has an outflow extent of only $\sim$\,13\,pc and maximum velocities of $\sim$\,100\,{\kms}.

Going to higher stellar masses, in the central galaxy of the Phoenix cluster (Phoenix A, with $M_*$ estimated by \citet{mcdonald+14} to be ${\sim}\,10^{11.9}$\,{\msun}), the {\coIII} intensity map displays a similar pattern to that observed in our $M_0$ map (Figs. \ref{fig:fig0} and \ref{fig:map_ngc3281}). Both present an S-shape morphology, and, in the case of Phoenix A, the {\co} surrounds the edges of a bipolar X-ray emission \citep{russel+17}. However, 
in the case of the Phoenix galaxy, the {\oiii} seems to extend along the {\co} emission \citep{mcdonald+14}, 
and a radio jet driven bubble is argued to be responsible for the X-ray cavities and to drive the molecular gas outflow encasing these cavities \citep{russel+17}. 
Other examples of the molecular gas avoiding/surrounding radio lobes includes the young radio galaxy PKS 0023-26 \citep{morganti+21} and the 3C 305 \citep{morganti+23}.

Other Seyfert galaxies present similar features in the hot molecular gas, measured via H$_2$ lines in the near-infrared. The hot H$_2$ has been detected in regions surrounding the ionisation axis, sometimes even showing an ``hour-glass'' shape, in objects like  NGC\,4151 \citep{may+20}, MRK\,573 \citep{fischer+17} and NGC\,1068 \citep{may+17},  
with the latter also showing a similar distribution in {\coIII} \citep{garcia-burillo+16}. All these objects also present radio jets along the ionisation axis, and most of them show signs of hot molecular gas outflows.

\section{Conclusions}\label{sec:conclusions}

We analysed the cold molecular gas distribution and kinematics of the Seyfert\,2 galaxy NGC\,3281 using observations of the {\coII} line. For such, we employed ALMA observations with an angular resolution of 0.5{\arcsec} ($\sim$\,100\,pc). The main results are:

\begin{itemize}
    \item We found an anti-correlation between the spatial distribution of the {\co} molecular gas emission -- mostly seen in the galaxy plane -- and that of the ionised gas, that shows a biconical geometry previously observed in the optical {\oiii}. The total molecular gas mass seen in emission is $M_{\mol,\EM}$\,=\,($1.5\pm0.9){\times}10^{8}$\,{\msun}. 
    \item Although most {\co} emission comes from the galaxy plane, we found it also surrounding the bipolar cones. This is supported by the $M_0$ moment, that shows CO emission extending upwards and downward from the plane. Its distribution suggests that part of the molecular gas is leaving the galaxy plane in the nuclear region but is being destroyed along the AGN ionisation axis.
    \item The {\co} kinematics is dominated by rotation in the galaxy plane, as shown by the fit of a 3D model to the {\coII} data cube. The  mean inclination of the disk and position angle of its axis were found to be $i_d$\,=\,139{\degree} and PA\,=\,73{\degree}, with the systemic velocity corresponding to a redshift of z\,=\,0.01124.
    \item From the residuals between the data and the rotation model, we identified two regions where the {\co} is tracing molecular gas outflows. One above the disk, and another below, both at the edges of the {\oiii} cones. Depending on the assumed molecular outflow inclination ($i_{mol}$\,=\,139 or 109{\degree}), it reaches maximum velocities of 160\,--\,360\,{\kms}, extending up to  $R_{\out}^{\rm{max}}$\,$\sim$\,1.8\,--\,2.6\,kpc from the nucleus. 
    \item The corresponding molecular mass in outflow is $M_{\mol,\out}$\,=\,$(2.5\pm1.6){\times}10^{6}$\,{\msun}, and  represents  $\sim$\,1.7\,\% of the total emitting molecular gas mass emission $M_{\mol,\EM}$. If we assume that the gas inside the B$_1$ regions are also outflowing, this fraction becomes $M_{\mol,\out}/M_{\mol,\EM}$\,$\sim$\,2\,\%,.
    \item The mass outflow rate is $\dot{M}_{\out,\mol}$\,=\,0.12\,--\,0.72\,{\msunyr} (depending on $i_{\mol}$). These values correspond to the peak of the radial profiles of $\dot{M}_{\out,\mol}(r)$. 
    The corresponding molecular outflow power is  $\dot{E}_{\out,\mol}$\,=\,(0.045\,--\,1.6)\,${\times}\,10^{40}$\,{\ergs}, which translates to a molecular kinetic efficiency of $10^{-4}$\,--\,0.02\,\%. Including the contribution of the ionised outflow and assuming that only 20\% of the AGN feedback couples kinetically with the gas, the maximum value of the coupling efficiency of the AGN feedback with the surroundings only reaches $\sim$0.1\,\%.
    \item There are other two regions with distinct {\co} residuals, that do not follow a rotation pattern, in patches running parallel to the major axis. Both can be interpreted as molecular inflow in the disk, given the corresponding velocities and the correlation with dust. One of them, however, could also be interpreted as being in outflow in front of the {\oiii} cone. Nonetheless, the interpretation of molecular gas in these regions is uncertain, and we do not attempt to calculate the corresponding rates of inflow/outflow.
\end{itemize}

The CO outflow geometry in NGC\,3281 -- encasing the ionised gas, is similar to those seen in other sources like NGC\,256, Milky Way and Perseus\,A, where the molecular gas outflow surrounds the ionised gas cone and/or the X-ray/radio emission. These results suggest that such molecular outflows can be common in active galaxies. But they can only be found with a careful analysis of the molecular gas distribution and kinematics, to separate gas rotating in the galaxy plane from that in outflow.

We also note that the low coupling efficiency and small/medium velocities indicate that the current feedback probably won't affect much the evolution of the host galaxy NGC\,3281, at least over larger scales. However, the expelling of molecular gas from inner regions will diminish the SFR over short periods of time, since it ends up removing a fraction of the fuel available to form new stars.

\section*{Acknowledgements}

This paper makes use of the following ALMA data: ADS/JAO.ALMA\#2018.1.00211.S. ALMA is a partnership of ESO (representing its member states), NSF (USA) and NINS (Japan), together with NRC (Canada), MOST and ASIAA (Taiwan), and KASI (Republic of Korea), in cooperation with the Republic of Chile. The Joint ALMA Observatory is operated by ESO, AUI/NRAO and NAOJ.

This study was financed in part by the Coordena{\c c}{\~a}o de Aperfei{\c c}oamento de Pessoal de N\'ivel Superior (CAPES-Brasil, 88887.478902/2020-00). R.A.R. acknowledges the support from Conselho Nacional de Desenvolvimento Cient\'ifico e Tecnol\'ogico (CNPq) and Funda\c c\~ao de Amparo \`a pesquisa do Estado do Rio Grande do Sul (FAPERGS).

 This work made use of {\tt Astropy}: a community-developed core Python package and an ecosystem of tools and resources for astronomy \citep{astropy:2022}. We also used the {\tt spectral-cube} package \citep{spectral-cube:2019} to manipulate the data cube, and {\tt Matplotlib} \citep{matplotlib} to generate the figures. 
 

\section*{Data Availability}
The data used in this paper are publicy available at the ALMA Science Archive under the program code 2018.1.00211.S. Processed data can be shared on reasonable request to the corresponding author.




\bibliographystyle{mnras}
\bibliography{bib} 




\appendix

\section{Moments}\label{sec:ap-moments}

\begin{figure}
    \centering
	\includegraphics[width=0.85\columnwidth]{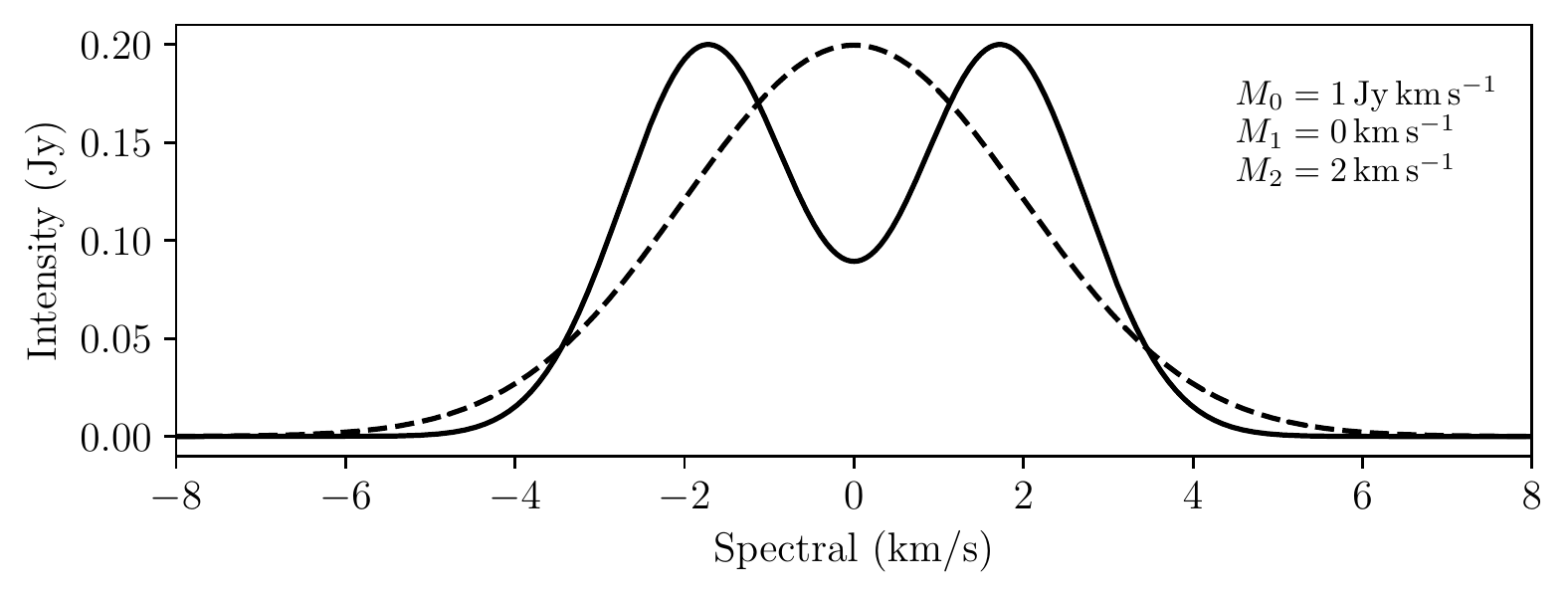}
    \caption{Two (synthetic) distinct emission-line profiles that have the same $M_0$, $M_1$ and $M_2$ moments values.}
    \label{fig:moment}
\end{figure}

For each spaxel the moments were calculated as:
\begin{eqnarray}
M_0&=&\sum S(\vel) \Dv\\
M_1&=&\frac{\sum \vel\, S(\vel) \Dv}{M_0}\\
M_2&=&\sqrt{\frac{\sum(\vel-M_1)^2  S(\vel) \Dv}{M_0}},
\end{eqnarray}
where $M_0$ is the total flux of the line profile, while $M_1$ and $M_2$ are the intensity weighted radial velocity and velocity dispersion (a measure of the internal disturbance of the gas), respectively. The flux density ($S(\vel)$) are in units of {\jybeam}, while the channel central velocity ($\vel$) and width ($\Dv$) are in {\kms}).
For a perfect Gaussian profile, $M_1 = v$ (the radial velocity) and $M_2 = \sigma$ (velocity dispersion). But we note that, in some cases, the moments can hide more complex profiles. An example is shown in Fig.\,\ref{fig:moment}, where one simple Gaussian and a double-peaked profile -- possibly denoting the presence of two kinematic components -- the same values for their moments $M_0$, $M_1$ and $M_2$. 

\section{Masks}\label{sec:ap-mask}

\begin{figure}
    \centering
	\includegraphics[width=1.\columnwidth]{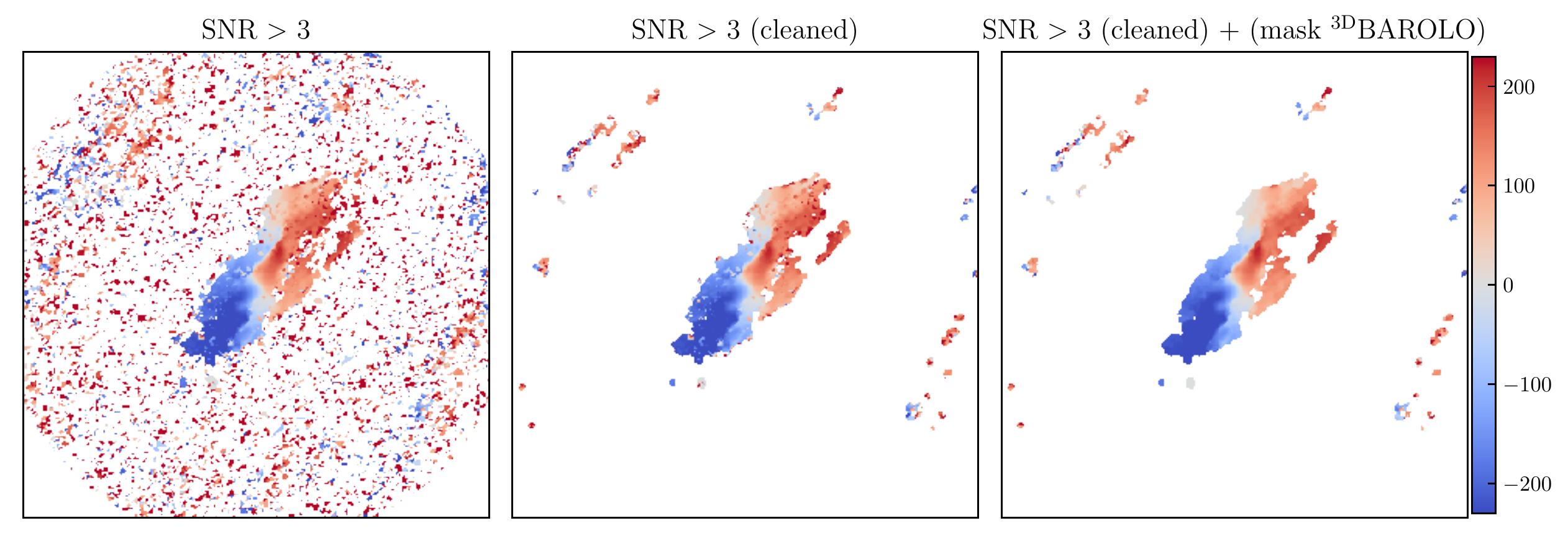}
    \caption{$M_{1}$ map calculated using different masks. Left: all spaxels whose spectra do not have flux density values above 3\,{\rms} are masked. 
    Center: the left 2D mask after being cleaned,} by using a cutting threshold applied to a smoothed version of the mask. Right: The combination of the mask from the middle with the 3D mask produced by {\barolo} that uses a similar threshold method applied to the cube (channel by channel).
    \label{fig:mask_ngc3281}
\end{figure}

To explain how the mask used to identify regions with dominating {\co} emission lines was created, in Fig.\,\ref{fig:mask_ngc3281}, we show how the $M_1$ map is affected when different masks are used.

First, we selected all spaxels that have at least one spectral value above 3\,{\rms},
generating a 2D mask from remaining ones (left panel of Fig.\,\ref{fig:mask_ngc3281}). Using only this criterion, a lot of spurious spaxels are not masked. 

To ``clean'' these isolated spaxels, we smoothed the mask  spatially with a 2D Gaussian with $\sigma$\,=\,1\,pix. This was done to the mask in a binary (0 and 1 values) form.
Then, any spaxels with resulting values lower than 0.5 were and masked. This process masks isolated group with less then 5 spaxels (area $\sim$\,8 times smaller then the beam one), and was repeated until a convergence was achieved. The middle panel shows the $M_1$ obtained with the resulting mask. 

However, since this is a 2D mask, spurious values along the spectral dimension can still affect the final moment values.
Therefore, we combine the 2D mask with the 3D one produced by {\barolo} using the ``SMOOTH'' option -- with the default parameters SCALEFACTOR\,=\,2 and BLANKCUT\,=\,3 -- that smooths each channel of the cube individually and masks them based on the resulting signal-to-noise ratio. The resulting 3D mask was combined with the 2D one, and used to obtain the final moment maps (see right panel of Fig.\,\ref{fig:mask_ngc3281}, and Figs.\,\ref{fig:map_ngc3281} and \ref{fig:barolo_ngc3281}). 

\section{Additional figures}\label{sec:ap-figures}

\begin{figure}
    \centering
	\includegraphics[width=1\columnwidth]{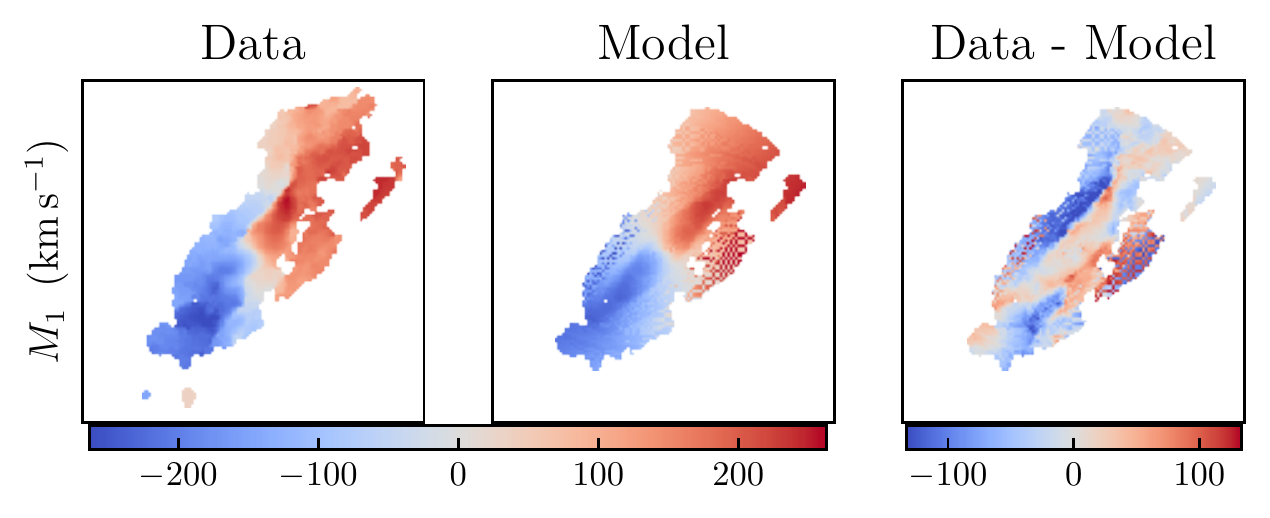}
    \caption{
        Fitting results of the 2D model \citep{begeman87} fitted to $M_1$ kinematics using {\barolo}. 
        The first, second and third columns correspond to $M_1$ data, the 2D model and the corresponding residuals, respectively.
}
    \label{fig:barolo2D}
\end{figure}

The Fig.\,\ref{fig:barolo2D} shows the results of the fit of a 2D model \citep{begeman87} on the first moment of {\coII} ($M_{1}$) .

\bsp	
\label{lastpage}
\end{document}